\PassOptionsToPackage{hyphens}{url}
\PassOptionsToPackage{dvipsnames,svgnames,x11names}{xcolor}

\documentclass[authoryear,3p,onecolumn]{elsarticle}

\usepackage[T1]{fontenc}
\usepackage[utf8]{inputenc}
\usepackage{textcomp}
\usepackage{lmodern}
\usepackage{xcolor}
\usepackage{amsmath,amssymb}

\usepackage{longtable,booktabs,array}
\usepackage{calc}
\usepackage{multirow}
\usepackage{siunitx}

\usepackage{graphicx}
\makeatletter
\def\fps@figure{htbp}
\makeatother

\usepackage{float}
\usepackage[table]{xcolor}

\usepackage{caption}
\usepackage{subcaption}

\usepackage{natbib}
\usepackage{bookmark}
\usepackage{xurl}

\providecommand{\tightlist}{%
  \setlength{\itemsep}{0pt}\setlength{\parskip}{0pt}}

\journal{in-progress}
\begin{document}

\begin{frontmatter}
\title{Code Contribution and Credit in Science}
\author[1]{Eva Maxfield Brown (0000-0003-2564-0373)%
\corref{cor1}%
}
 \ead{evamxb@uw.edu} 
\author[1]{Isaac Slaughter (0000-0002-1911-2374)%
}

\author[1]{Nicholas Weber (0000-0002-6008-3763)%
}

\affiliation[1]{organization={University of Washington Information
School},,postcodesep={}}

\cortext[cor1]{Corresponding author}

\begin{abstract}
Software development has become essential to scientific research, but
its relationship to traditional metrics of scholarly credit remains
poorly understood. We develop a dataset of approximately 140,000 paired
research articles and code repositories, and a predictive model that
matches research article authors with software repository developer
accounts. We use this dataset to investigate how software development
activities influence credit allocation in collaborative scientific
settings. Our findings reveal significant patterns distinguishing
software contributions from traditional authorship credit. We find that
$\sim$30\% of articles include non-author code
contributors---individuals who participated in software development but
received no authorship recognition. While code-contributing authors
provide a $\sim$4.2\% increase in article citations, this
effect becomes non-significant when controlling for domain, article
type, and open access status. First authors are significantly more
likely to be code contributors than other author positions. Notably, we
identify a negative relationship between coding frequency and scholarly
impact metrics. Authors who contribute code more frequently exhibit
progressively lower h-indices than non-coding colleagues, even when
controlling for publication count, author position, domain, and article
type. These results suggest a disconnect between software contributions
and credit, highlighting important implications for institutional reward
structures and science policy.
\end{abstract}


\end{frontmatter}

\section{Introduction}\label{introduction}

Recent advances in genomic sequencing, climate modeling, particle
physics, and neuroimaging have all required the development of novel
scientific software \citep{hocquet2024software}. In many ways, software
development has changed how scientific work is organized and performed,
but we lack large-scale quantitative evidence characterizing these
effects. This lack of evidence has important consequences. Scientific
institutions continue to rely heavily on bibliometric indicators to
evaluate research productivity \citep{haustein2014use}, but software is
rarely mentioned or cited in research publications
\citep{du2021softcite}. As a result, researchers who invest substantial
effort in building computational tools may face systematic disadvantages
in career advancement. We argue that understanding whether and how
software contributions translate into formal scientific credit is
important for designing equitable reward structures and evidence-based
science policy.

The challenge of measuring software's role in science stems, in part,
from the historical separation of code from publication. Scientific
articles and their metadata (e.g. author lists, citations, and
acknowledgments) can provide a valuable tool for estimating
productivity, tracing collaboration, or inferring impact
\citep{fortunato2018science}. However, software contributions often
remain invisible in these records, making it difficult to quantify their
relationship to traditional markers of scientific productivity
\citep{weber2014paratexts}. As a result, studies regarding the
maintainers and contributors of scientific software have relied
primarily on self-reports (surveys) \citep{Carver2022ASO} or participant
observation (ethnographies) \citep{paine2017has}, which provide valuable
insights but are limited in the scope and generalizability.

In the following paper, we overcome this challenge by constructing a
dataset (\texttt{rs-graph-v1}) of 138,596 research articles and their
associated code repositories. Using this data we then develop a machine
learning model to match article authors with repository contributor
accounts. By connecting papers to repositories, and authors to
developers, we are able to observe both the presence and effect of code
contributions across different authorship positions and examine how
coding activity relates to both article-level citations and career-level
productivity metrics (e.g., h-index). This approach enables us to
address fundamental questions about the scientific reward system: How do
software contributions correspond to authorship positions? Does coding
effort translate into citation impact? How do patterns of code
contribution relate to long-term career trajectories?

In analyzing a filtered subset of the \texttt{rs-graph-v1} dataset, we
identify three patterns that distinguish code contributions from other
forms of scientific recognition. First, we show that 28.6\% (n=5,694) of
articles have non-author code-contributors, or individuals who may have
helped create the software but received no formal authorship credit. We
also show that code-contributing authors are associated with a modest
increase in article-level impact metrics (on average, a 4.2\% increase
in citations per code-contributing author), but these effects become
statistically non-significant when controlling for domain, article type,
and open access status. Second, we find that first authors are
significantly more likely to be code contributors than other author
positions across all conditions tested. Third, we find a negative
relationship between coding frequency and scholarly impact: Authors who
contribute code more frequently show progressively lower h-indices than
their non-coding peers, a pattern that persists even when controlling
for publication count, and author's most common author position (first,
middle, or last), domain (Physical Sciences, Health Sciences, Life
Sciences, or Social Sciences), and article type (preprint, research
article, or software article)\footnote{We use the phrase ``most common''
  to mean most frequent, and in the case of a tie, their most recent.
  That is, if an author has been first author on four publications in
  our dataset, middle author on two publications, and has never been
  last author, they are considered a ``first author'' for the purposes
  of our analysis. We discuss the limitations of this approach in the
  Section~\ref{sec-discussion}.}.

The remainder of this paper proceeds as follows: First, we review
related work regarding software development and the recognition of code
contributors in scientific credit systems. In doing so, we motivate
three specific hypotheses that guide our investigation. Next, we provide
an overview of our data and methods, describing how we linked articles
to repositories, trained and evaluated a predictive model for entity
matching, and applied this model across each article-repository pair. We
then present our analysis, focusing on article-level dynamics before
moving to individual-level (career) patterns, formally accepting or
rejecting each hypothesis based on our findings.

\section{Background}\label{sec-background}

The proper allocation of credit has been a longstanding topic in social
studies of science. Merton's description of a ``Matthew Effect'' is
perhaps the most famous account of a cumulative advantage in publishing,
where established scientists receive more attention for work of similar
importance or value than their less established peers
\citep{merton1968matthew}. Contemporary studies of science continue to
demonstrate important aspects of the dilemma. Dashun Wang et al. show
that random career outcomes and the ``hot streak'' phenomenon suggest
that cumulative advantage and early career success may be less
predictive of long-term impact than traditionally assumed, complicating
how we understand the accrual of scientific credit over time
\citep{wang2013quantifying, liu2018hot}. Separately, work on team
science demonstrates that larger teams receive disproportionate credit
compared to smaller teams producing equally impactful work, further
illustrating how credit allocation deviates from underlying
contributions \citep{wu2019large}.

In quantitative work, the attribution of credit is most often
established using proxies in bibliographic data like author order
\citep{zuckerman1968patterns, sarsons2017recognition, sauermann2017authorship},
which is an imperfect means of assigning credit, and doing so may
exacerbate existing inequalities \citep{west2013role}. More recent work
developing algorithms to assign credit \citep{shen2014collective} and
formally document contributor roles \citep{allen2014publishing} are
aimed at trying to improve upon author order as the status quo mechanism
for assigning credit in collaborative settings. Further, the use of
contributor statements as data have provided important insights into the
actual division of labor in scientific collaborations and revealed
systematic patterns in how different types of contributions are valued
and recognized across fields \citep{lu2020co, lu2022contributorship}.

Among the contributions systematically undervalued in collaborative
science are technical and infrastructural contributions, particularly
software development. While contributor role taxonomies have begun to
make visible the diverse forms of labor that constitute scientific work,
these systems still struggle to adequately recognize the significance of
technical contributions that enable research, but may not fit
traditional notions of intellectual authorship
\citep{Larivire2020InvestigatingTD}. Even less is known about the
broader patterns of attention that software-producing research teams
receive, and under what conditions technical contributions achieve
visibility in an attention economy driven by citations. We argue that
understanding these patterns requires examining how team composition and
software contributions relate to scientific attention at scale.

Next, we build upon these general findings to motivate three specific
hypotheses about code contributions and credit that we then test using
the \texttt{rs-graph-v1} dataset.

\subsection{H1: Research Team Composition and Scientific
Attention}\label{h1-research-team-composition-and-scientific-attention}

In collaborative settings, experimental and theoretical research often
receives more citations than methods-focused contributions, with the
exception being when methods represent foundational shifts in a field
\citep{Aksnes2006CitationRA, LIU2023101432, Chen2024ExploringSC}.
Software is often positioned as a methodological contribution and, as a
result, can be found in some of the highest-cited papers of the 21st
century
\citep{jin2015significance, hampton2013big, hasselbring2024researchsoftwarecategories}.

Prior work also establishes a positive relationship between team size
and citation count, where larger and more diverse teams produce
higher-impact scientific work
\citep{Franceschet2010TheEO, Larivire2014TeamSM, yoo2024interaction}.
Research in empirical software studies similarly finds that larger
development teams tend to create more reliable software with fewer
defects \citep[\citet{herbsleb2003empirical}]{Wyss2023NothingBut},
though this comes at the expense of slower development cycles. These
findings suggest that team size may be particularly important in
scientific software development, where technical robustness and
reproducibility remain gold standards \citep{milewicz2018talk}.

We argue that the unique characteristics of scientific software
development --- including implementing novel algorithms, requiring deep
domain knowledge, and an increased emphasis on reproducibility
\citep{muna2016astropyproblem, Howison2013IncentivesAI} --- make team
composition especially important for understanding credit allocation.
Software development in organized teams may enhance scientific impact
through multiple mechanisms: teams can produce more robust and
generalizable software tools for methodological contributions while
enabling more sophisticated computational analyses and larger-scale data
processing for experimental work.

Given these patterns in team dynamics, software development practices,
and citations in collaborative research, we propose that:

\emph{H1: The number of individuals contributing code to a publication's
associated repository positively correlates with the article's citation
count.}

\subsection{H2: Author Roles and Technical
Contributions}\label{h2-author-roles-and-technical-contributions}

Author positions in scientific publications signal specific roles and
responsibilities \citep{shen2014collective}, a relationship that can be
contemporarily studied through contribution role taxonomies like CRediT
\citep{lariviere2016contributorship}. In this literature, analyses of
contribution patterns consistently show that software development, data
analysis, and visualization tasks typically fall to first authors
\citep{lariviere2016contributorship, Jnior2016PatternsOA, Larivire2020InvestigatingTD, sauermann2017authorship}.
While these contribution patterns are often inferred from self-reported
roles (e.g., CRediT) or disciplinary conventions, the linking of
publications to version-controlled software repositories provides a
novel window into the actual coding contributions associated with
scientific papers. Given the established tendency for first authors to
undertake primary research execution, including technical development,
we anticipate that this pattern will be directly observable via the
commit histories of scientific software repositories. Given prior
findings about the distribution of technical responsibilities within
research teams, we expect these repository records to reflect similar
patterns of engagement with software development:

\emph{H2a: First authors have higher code contribution rates than
authors in other positions.}

Furthermore, studies that use contribution taxonomies like CRediT reveal
that first authors and corresponding authors, while occasionally the
same individual, often take on distinct responsibilities
\citep{birnbaum2023should}. Corresponding authors traditionally bear
responsibility for the integrity, archiving, and long-term availability
of research artifacts and underlying data
\citep{teunis2015corresponding, sauermann2017authorship, da2013corresponding, BHANDARI20141049}.
As software becomes an increasingly critical, complex, and integral
research artifact in many disciplines, it logically follows that
stewardship of these software components should align with the
responsibilities of being a corresponding author. Given prior findings
about the distribution of responsibilities within research teams and the
expectation of stewardship of created artifacts, we expect repository
records to reflect similar patterns of engagement with software
development:

\emph{H2b: Corresponding authors have higher code contribution rates
than non-corresponding authors.}

\subsection{H3: Code Contribution and Individual Scientific
Impact}\label{h3-code-contribution-and-individual-scientific-impact}

Despite the increasingly central role of software in science,
researchers who dedicate significant effort to its development often
face systemic hurdles in receiving formal scientific credit. Their
contributions may be relegated to acknowledgment sections rather than
rewarded with authorship credit \citep{weber2014paratexts}. Further, the
scholarly practice of software citation remains inconsistent, frequently
undervaluing crucial maintenance and extension work
\citep{Carver2022ASO, olivier_philippe_2019_2585783, Lamprecht2020TowardsFP, Katz2020RecognizingTV, Smith2016SoftwareCP}.
The h-index, a widely used proxy for an individual's cumulative
scientific impact, is derived from an individual's record of formally
authored publications and the citations these publications receive
\citep{hirsch2005index}. Consequently, if substantial time and
intellectual effort are invested in software development that does not
consistently translate into formal authorship on associated research
papers, or if the software outputs themselves are not robustly and
formally cited in a way that accrues to the individual developer, then
the primary activities that build a researcher's h-index are effectively
diminished or bypassed.

This creates a structural misalignment where contributions essential for
scientific advancement (i.e., software development and maintenance) may
not adequately capture and could even detract from time spent on
activities that bolster traditional bibliometric indicators of
individual success. While collaborative software development may yield
short-term benefits through increased citations on individual papers (as
suggested by H1), researchers specializing in code development may face
long-term career disadvantages as their expertise becomes increasingly
divorced from traditional publication pathways that drive academic
recognition and advancement. Based on these challenges in the
recognition and citation of software contributions and their potential
impact on h-index accumulation, we hypothesize:

\emph{H3: The frequency with which individual researchers contribute
code to their research projects negatively correlates with their
h-index.}

\section{Data and Methods}\label{data-and-methods}

We examine the relationship between software contributions and
scientific credit through a three-step process: (1) building a dataset
of linked scientific articles and code repositories; (2) developing a
predictive model to match article authors with developer accounts; and,
(3) analyzing patterns in these relationships.

\begin{figure}

\centering{

\includegraphics[width=3.85in,height=7in]{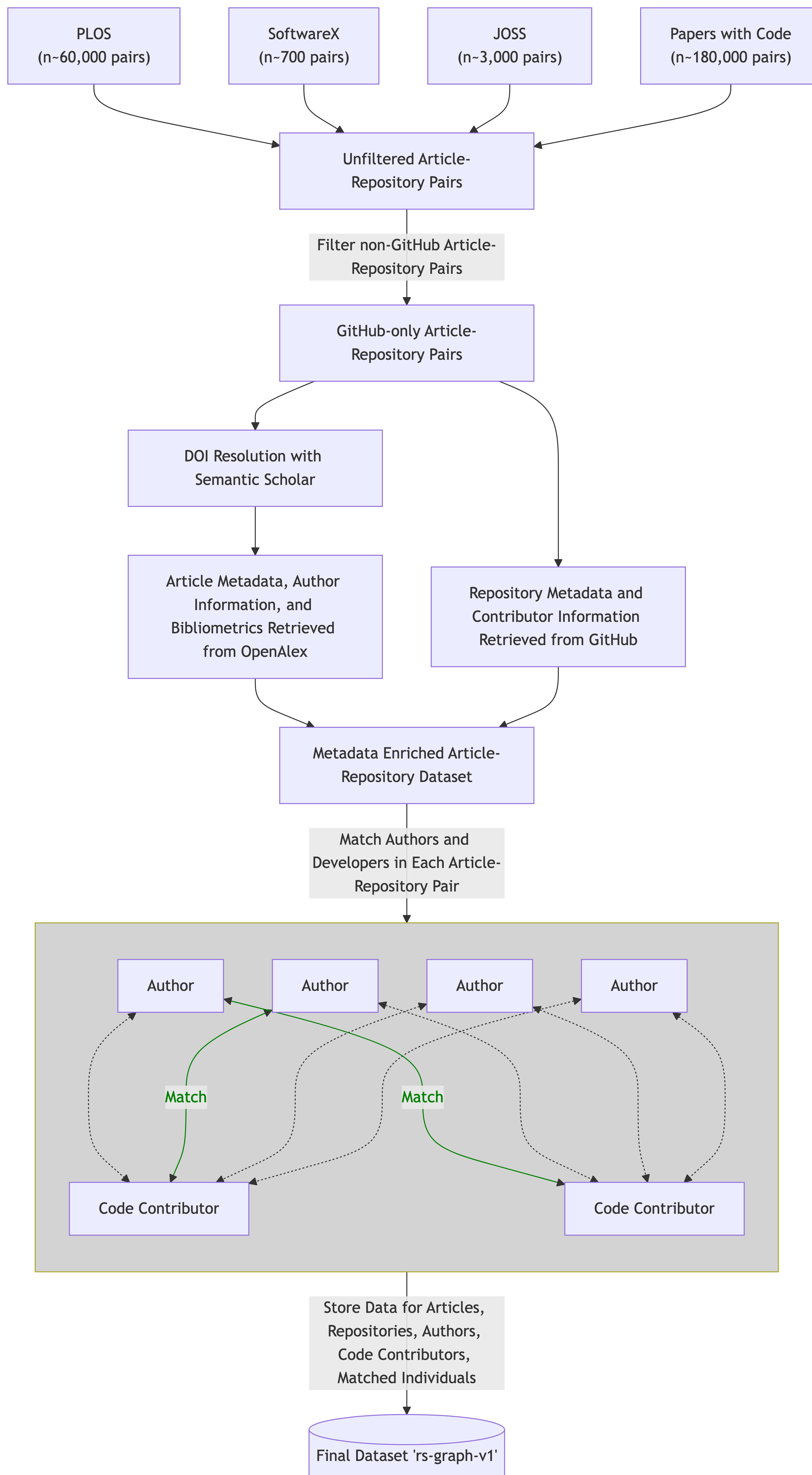}

}

\caption{\label{fig-workflow-diagram}Data Gathering and Processing
Workflow. We begin by gathering article-repository pairs from PLOS,
JOSS, SoftwareX, and Papers with Code, retaining only pairs where the
source code repository is hosted on GitHub. For each article-repository
pair, we retrieve the DOI for the most recent version of the article via
the Semantic Scholar API, then obtain the article's metadata, author
information, and bibliometric data from OpenAlex. In parallel, we
collect the source code repository's metadata and contributor
information using the GitHub API. After retrieving both article and
repository information, we apply our trained author-developer matching
model across all possible combinations of article authors and repository
contributors for each article-repository pair. Finally, we store all
retrieved information and the matched author-developer pairs in a SQLite
database. Article-repository pair counts are approximate because there
are no snapshots of these databases at a single point in time. The
estimates are based on counts from each data source taken in October
2025.}

\end{figure}

Our dataset integrates article-repository pairs from four sources, each
with explicit mechanisms for code sharing: the Journal of Open Source
Software (JOSS) and SoftwareX require code repositories as part of
publication, Papers with Code directly connects preprints from ArXiv
with software implementations, and Public Library of Science (PLOS)
articles include mandatory data and code availability statements that we
mined for repository links. We focused exclusively on GitHub-hosted
repositories, which represent the predominant platform for scientific
software sharing \citep{Cao2023TheRO, escamilla2022riseofgithub}. For
each article in our corpus, we resolved the source DOI via Semantic
Scholar to ensure we captured its latest version and then extracted
publication metadata and author metrics through OpenAlex. Finally, we
collected information about each code repository and the repository's
contributors via the GitHub API. A data collection and processing
workflow diagram is available in Figure~\ref{fig-workflow-diagram}. All
data was collected between October and November 2024\footnote{The
  Journal of Open Source Software (JOSS) is available at
  \url{https://joss.theoj.org/}. SoftwareX articles are available at
  \url{https://www.sciencedirect.com/journal/softwarex/} and SoftwareX
  repositories are available at
  \url{https://github.com/ElsevierSoftwareX/}. Public Library of Science
  (PLOS) is available at \url{https://plos.org/} and article data is
  available via the \texttt{allofplos} Python package
  (\url{https://github.com/PLOS/allofplos/}). Papers with Code is
  available at \url{https://paperswithcode.com/} and data for links
  between papers and code is available at
  \url{https://paperswithcode.com/about/}. Documentation for the
  Semantic Scholar API is available at
  \url{https://api.semanticscholar.org/}, documentation for the OpenAlex
  API is available at \url{https://docs.openalex.org/}, and
  documentation for the GitHub API is available at
  \url{https://docs.github.com/en/rest/}.}.

Articles collected from JOSS and SoftwareX were labeled as ``software
articles,'' articles obtained from PLOS were labeled as ``research
articles,'' and articles from Papers with Code were labeled as either
``preprint'' or ``research article'' based on their document type from
OpenAlex. While Papers with Code data is largely tied to ArXiv
preprints, because of our DOI resolution process, we were able to
identify a subset of these articles that had been published in a journal
and were thus labeled as ``research articles'' by OpenAlex\footnote{Papers
  with Code (arXiv) preprints likely include some number of software
  articles but without a clear and consistent mechanism to identify
  these articles, but we are unable to label them as such. We discuss
  this limitation further in Section~\ref{sec-discussion}. Further,
  while some PLOS, JOSS, and SoftwareX articles may also have a
  preprinted version, we select and analyze the published version of the
  article when one is available at the time of data processing. This is
  made possible via the DOI resolution process using Semantic Scholar.}.
Article domain classifications (e.g., Health Sciences, Life Sciences,
Physical Sciences, and Social Sciences) were obtained from OpenAlex.
While each article can be associated to multiple ``topics'', we utilize
the primary topic and its associated domain for our analyses\footnote{OpenAlex
  documentation for concepts, topics, fields, and domains is available
  at
  \url{https://help.openalex.org/hc/en-us/articles/24736129405719-Topics/}}.

To match article authors with repository developer accounts, we
developed a machine-learning approach using transformer-based
architectures. Specifically, we use transformers to overcome entity
matching challenges such as when exact name or email matching is
insufficient due to formatting variations and incomplete information
(e.g., ``J. Doe'' vs. ``Jane Doe'' in publications or use of
institutional versus personal email addresses). When exact matching
fails, there is typically high semantic overlap between an author's name
and developer account details (i.e., username, name, and email) that
transformer models can leverage. We created a gold-standard dataset of
2,999 annotated author-developer account pairs from JOSS publications,
where two independent reviewers classified each pair as matching or
non-matching. After systematic evaluation of three transformer
architectures with various feature combinations, our optimal model
(fine-tuned DeBERTa-v3-base including developer account username and
display name in the training data) achieved a binary F1 score of 0.944,
with 0.938 precision and 0.95 recall (positive = ``match'')\footnote{A
  detailed comparison of models and feature sets, and an evaluation of
  model performance across non-JOSS author-developer pairs is made
  available in the Appendix.}. Applying our model across all
article-repository pairs yielded a large-scale dataset linking
scientific authors and code contributors.

Our complete dataset, named the ``Research Software Graph''
(\texttt{rs-graph-v1}), contains 163,292 article-repository pairs.
However, 24,696 article-repository pairs form many-to-many relationships
in which the same article is linked to multiple repositories or the same
repository is linked to multiple articles. The median number of
repositories per article is 1 (95th percentile: 1, 97th percentile: 2,
99th percentile: 2). The median number of articles per repository is 1
(95th percentile: 1, 97th percentile: 2, 99th percentile: 2).

To ensure that our analysis focuses on clear, unambiguous relationships
between articles and repositories, prior to analysis we utilize a
partial \texttt{rs-graph-v1} dataset which filters out any articles or
repositories associated to multiple article-repository pairs. This
filtering step removes 24,696 article-repository pairs. As shown in
Table~\ref{tbl-rs-graph-overall-counts}, the complete one-to-one
article-repository dataset contains 138,596 unique article-repository
pairs spanning multiple domains and publication types (2,336 unique
articles, repositories, and pairs from JOSS, 6,090 from PLOS, 129,615
from Papers with Code, and 555 from SoftwareX). Additionally, the
one-to-one dataset includes information for 295,806 distinct authors and
152,170 developer accounts. The one-to-one dataset includes 90,086
author-developer account relationships, creating a unique resource for
investigating code contribution patterns in scientific teams. This
dataset enables systematic examination of how software development work
relates to scientific recognition and career trajectories. The complete
\texttt{rs-graph-v1} dataset (\url{https://doi.org/10.7910/DVN/KPYVI1}),
the trained model
(\url{https://huggingface.co/evamxb/dev-author-em-clf}), and a
supporting inference library for using the model:
\href{https://github.com/evamaxfield/sci-soft-models}{sci-soft-models},
are made publicly available to support further research.

\begin{table}

\caption{\label{tbl-rs-graph-overall-counts}Counts of Article-Repository Pairs, Authors, and Developers by Data Sources, Domains, Document Types, and Access Status in the Complete, Unfiltered Dataset.}

\centering{

    \centering
    \small

    \begin{tabular}{llrrr}
    \toprule
    \textbf{Category} & \textbf{Subset} & \textbf{Article-Repository Pairs} & \textbf{Authors} & \textbf{Developers} \\
    \midrule
        \multirow{4}{*}{\textbf{By Domain}} & \cellcolor{gray!10}Health Sciences & \cellcolor{gray!10}5,172 & \cellcolor{gray!10}25,979 & \cellcolor{gray!10}7,248 \\
    & Life Sciences & 7,729 & 31,649 & 12,150 \\
    & \cellcolor{gray!10}Physical Sciences & \cellcolor{gray!10}116,600 & \cellcolor{gray!10}240,545 & \cellcolor{gray!10}130,592 \\
    & Social Sciences & 8,838 & 29,269 & 14,043 \\\midrule
    \multirow{3}{*}{\textbf{By Document Type}} & \cellcolor{gray!10}preprint & \cellcolor{gray!10}72,177 & \cellcolor{gray!10}170,301 & \cellcolor{gray!10}87,311 \\
    & research article & 63,528 & 173,183 & 78,935 \\
    & \cellcolor{gray!10}software article & \cellcolor{gray!10}2,891 & \cellcolor{gray!10}9,294 & \cellcolor{gray!10}12,868 \\\midrule
    \multirow{2}{*}{\textbf{By Access Status}} & \cellcolor{gray!10}Closed & \cellcolor{gray!10}5,740 & \cellcolor{gray!10}23,668 & \cellcolor{gray!10}9,352 \\
    & Open & 132,856 & 286,874 & 147,831 \\\midrule
    \multirow{4}{*}{\textbf{By Data Source}} & \cellcolor{gray!10}joss & \cellcolor{gray!10}2,336 & \cellcolor{gray!10}7,105 & \cellcolor{gray!10}11,362 \\
    & plos & 6,090 & 30,233 & 8,784 \\
    & \cellcolor{gray!10}pwc & \cellcolor{gray!10}129,615 & \cellcolor{gray!10}262,889 & \cellcolor{gray!10}134,926 \\
    & softwarex & 555 & 2,244 & 1,628 \\\midrule
    \textbf{Total} & & \textbf{138,596} & \textbf{295,806} & \textbf{152,170} \\
    \bottomrule
    \end{tabular}

}

\end{table}%

\section{Analysis of Code Contributor Authorship and Development
Dynamics of Research
Teams}\label{analysis-of-code-contributor-authorship-and-development-dynamics-of-research-teams}

\subsection{Software Development Dynamics Within Research
Teams}\label{software-development-dynamics-within-research-teams}

\begin{figure}

\centering{

\includegraphics[width=\textwidth]{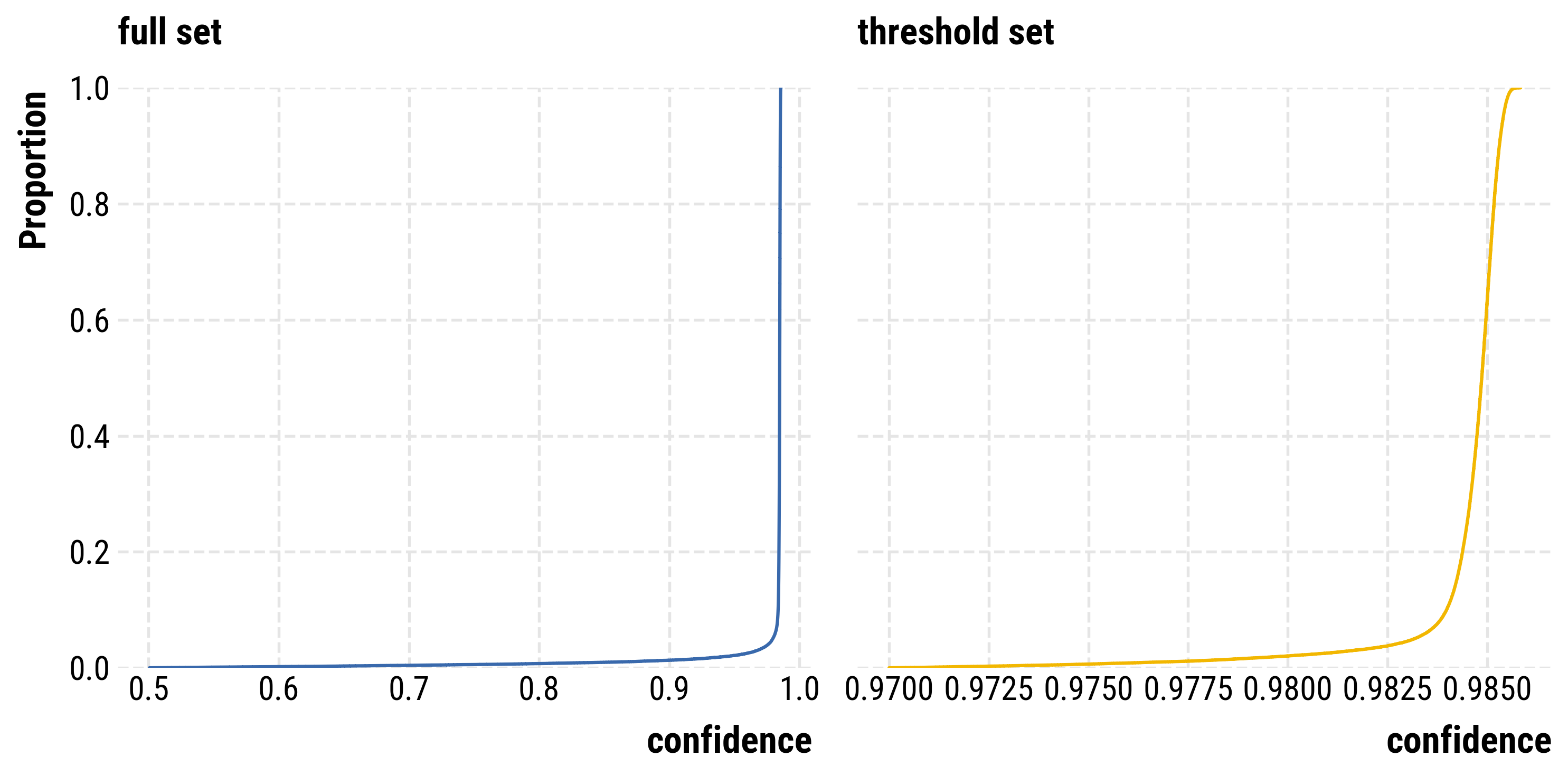}

}

\caption{\label{fig-dist-of-author-dev-pred-confidence}Distribution of
author-developer-account prediction confidence scores. The plot on the
left shows the distribution of all prediction confidence scores. The
plot on the right shows the distribution of prediction confidence scores
for author-developer-account pairs with a confidence score greater than
or equal to 0.97.}

\end{figure}

Understanding the composition and dynamics of software development teams
provides essential context for analyzing how code contributions relate
to scientific recognition and impact. To ensure reliable analysis, we
focus on a subset of our article-repository pairs that meet several
filtering conditions. First, to ensure that we aren't analyzing ``data
repositories'' (i.e. GitHub repositories which only store CSV, Parquet,
or other dataset related file formats), we filter out any
article-repository pairs that don't have any programming language
files\footnote{We use the list of programming languages used by the
  GitHub platform via the
  \href{https://github.com/github-linguist/linguist}{Linguist software},
  and the GitHub API to retrieve the bytes of code per-language within
  each repository. We remove article-repository pairs which have a
  non-zero number of bytes of code for programming language files.}.
Next, to ensure that the research has received a basic level of
engagement from the scientific community, we remove any
article-repository pairs which do not have at least one citation. We
then require that repository commit activity stop before 90 days past
the article publication date. Disallowing long-term projects ensures we
do not include projects that may add additional code contributors later
while still allowing a grace period during which developers can update
repositories with additional documentation and publication information.
We then subset the data to only include article-repository pairs with
research teams of typical size by removing those with fewer than three
authors and more than 11 authors, the 97th percentile for research team
size. Finally, we filter out any predicted author-developer pairs with a
confidence score of less than 0.97 in order to improve robustness of our
downstream analysis\footnote{Figure~\ref{fig-dist-of-author-dev-pred-confidence}
  shows the distribution of predictive model confidence scores for
  author-developer pairs to justify this threshold. We chose the 0.97
  threshold to ensure that we only include high-confidence matches while
  retaining a large proportion of the data (87,175 author-developer
  pairs) as only 2,911 author-developer-account pairs have a confidence
  less than 0.97 in the whole unfiltered dataset.}. This filtering
process results in a dataset of 19,900 article-repository pairs. A table
with the counts of article-repository pairs, authors, and developers by
data sources, domains, document types, and access status for this
filtered dataset is shown in
Table~\ref{tbl-team-comp-no-push-after-pub-counts}.

\begin{figure}

\centering{

\includegraphics[width=\textwidth]{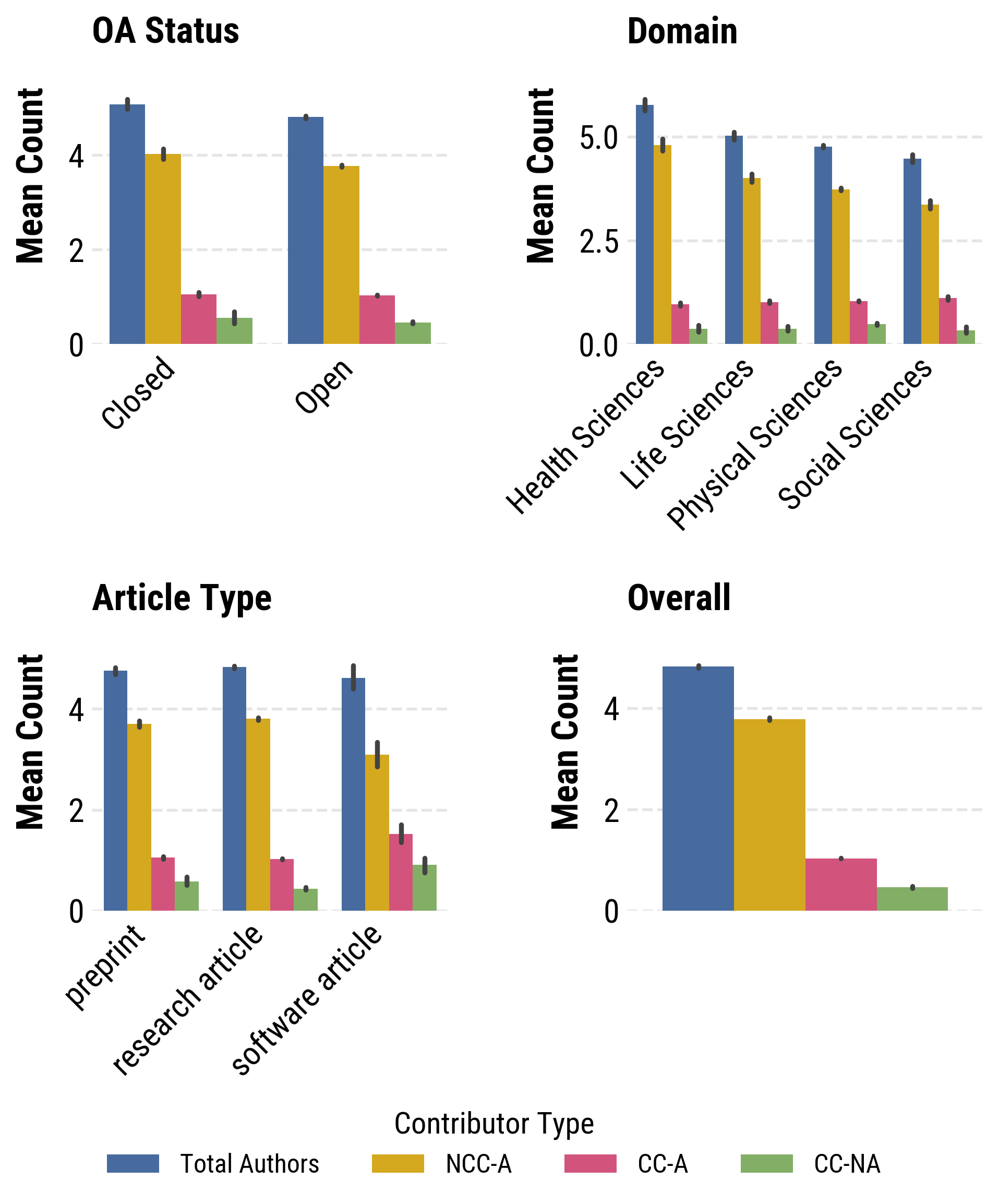}

}

\caption{\label{fig-team-composition}Mean of Non-Code-Contributing
Authors (NCC-A), Code-Contributing Authors (CC-A), and Code-Contributing
Non-Authors (CC-NA) Research Team Members by Domain, Article Type, and
Open Access Status. Only includes research teams from article-repository
pairs with repositories that have programming language files, a most
recent commit no later than 90 days after publication, and excludes
research teams in the top 3\% of total author sizes.}

\end{figure}

Within this filtered dataset, we categorized individuals into three
groups: code-contributing authors (CC-A) who both authored papers and
committed code to associated repositories\footnote{We define
  code-contributing authors as those who have at least one code commit
  to the repository associated with the article-repository pair.},
non-code-contributing authors (NCC-A) who authored papers but showed no
evidence of code contributions, and code-contributing non-authors
(CC-NA) who contributed code but received no authorship recognition.
This categorization revealed that papers in our dataset typically have
4.8 $\pm$ 1.8 total authors, with 1.0 $\pm$ 0.7 code-contributing authors and
3.8 $\pm$ 1.9 non-code-contributing authors. Beyond the author list, papers
averaged 0.5 $\pm$ 1.8 code-contributing non-authors.
Figure~\ref{fig-team-composition} details these distributions by domain,
article type, open access status, and overall across the entire filtered
analysis set\footnote{A table with the means and standard deviations for
  each distribution and subset of team composition is provided in the
  Appendix in Table~\ref{tbl-team-composition-counts}.}.

Perhaps most striking is our finding that 5,694 papers
($\sim$28.6\% of our sample) have at least one code contributor
who was not matched to any article author. Further, investigating this
subset of papers, we find an average of 1.6 $\pm$ 3.0 unmatched code
contributors per paper. We constructed a random sample of 200
code-contributing non-authors (see
Section~\ref{sec-appendix-additional-cc-na-analysis} for complete
methodology) in order to better understand these individuals and their
contributions. We characterize this subset in three ways:
$\sim$39\% (n=78) represent true non-authors (individuals who
likely should not be matched to an author), $\sim$30.5\% (n=61)
appear to be missed classifications (individuals who likely should have
been matched to authors), and the remaining $\sim$30.5\% (n=61)
were unclear due to limited profile information\footnote{While
  annotation for the model training and evaluation set was done using
  primarily simple text information (e.g., usernames, emails, etc.),
  annotation conducted during the qualitative analysis of
  code-contributing non-authors included a more in-depth review of
  public profiles including any linked websites or social media
  profiles. This was done to gain a true best estimate for the ``true'',
  ``missed'', and ``unclear'' classifications. Additional qualitative
  error analysis was conducted across the set of 61 likely authors and
  our findings are available in the Appendix
  (Section~\ref{sec-appendix-cc-na-error-analysis}).}.

Across all groups, the majority of the code-contributing non-authors
made code changes rather than just documentation updates. Among true
non-authors, $\sim$77.6\% (n=59) contributed code, with the top
25th percentile of these contributors contributing $\sim$10.7\%
of total repository commits and $\sim$14.4\% of absolute code
changes. The unclear cases showed substantially higher contribution
levels--- many were repository owners with extensive engagement. Yet,
even among the non-owners, the median contributor accounted for
$\sim$34.6\% of repository commits and $\sim$12.7\% of
absolute changes. Due to the limited size of our sample (n=200
code-contributing non-authors), we do not generalize these findings to
the entire population but note that they describe the diverse nature of
code contribution patterns from potentially unacknowledged contributors.

These findings reveal a more complex dynamic between software
development and authorship recognition than previously documented. While
our finding that each paper averages only a single code-contributing
author aligns with previous research showing technical tasks typically
fall to first authors \citep{Larivire2020InvestigatingTD}, the
substantial contributions made by unrecognized contributors suggest
systematic gaps in how scientific software development work is credited.

\begin{figure}

\centering{

\includegraphics[width=\textwidth]{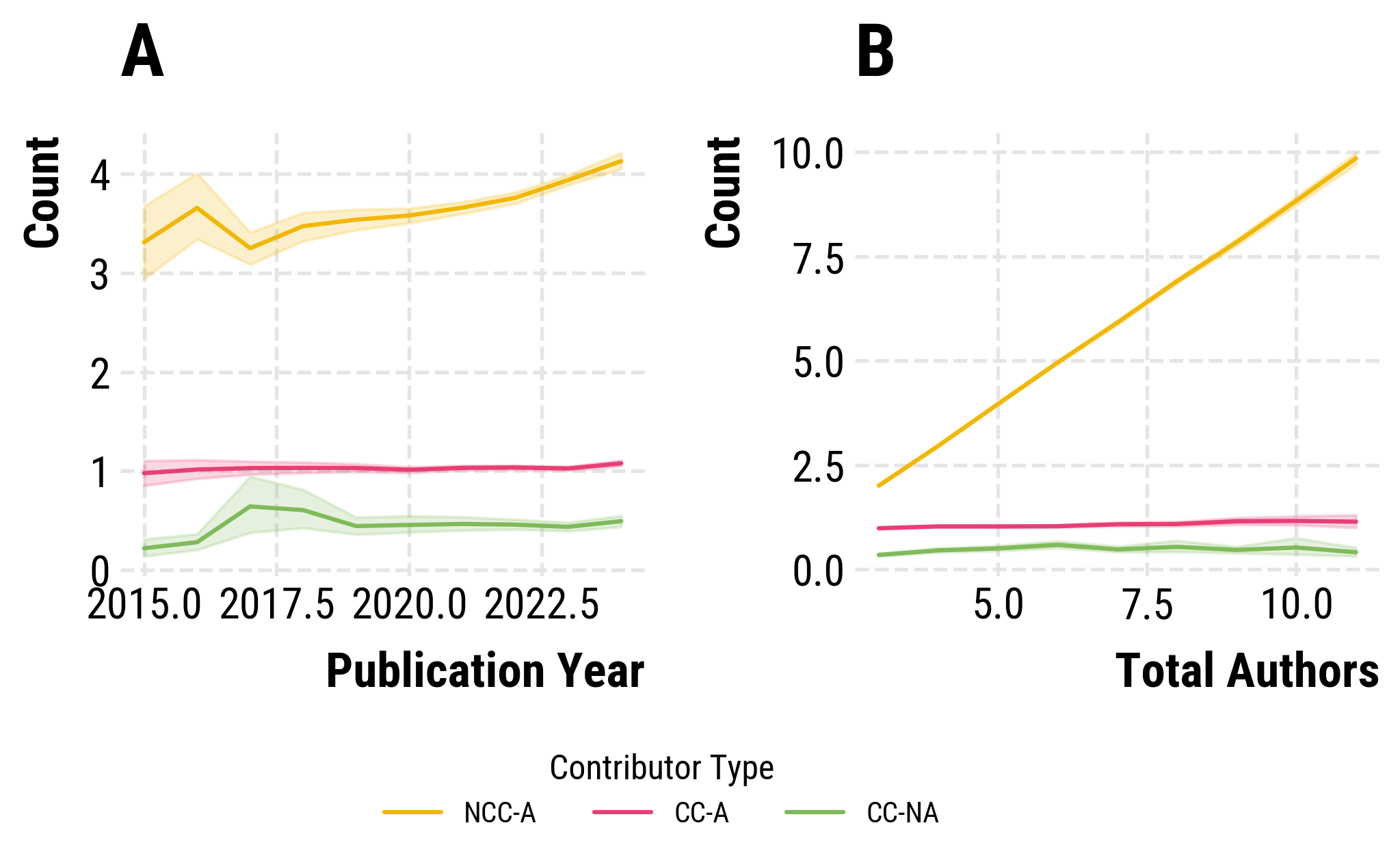}

}

\caption{\label{fig-contributor-type-by-time-and-size}Average number of
contributors per article, by contribution type: Non-Code-Contributing
Authors (NCC-A), Code-Contributing Authors (CC-A), and Code-Contributing
Non-Authors (CC-NA). Split into, A) the year the article was published
and B) the total number of authors included in the article. Only
includes research teams from article-repository pairs with repositories
that have programming language files, a most recent commit no later than
90 days after publication, and excludes research teams in the top 3\% of
total author sizes for publication years with 50 or more articles.
Shaded areas show the 95\% confidence interval for the mean.}

\end{figure}

When examining these patterns over time and across different team sizes
(Figure~\ref{fig-contributor-type-by-time-and-size}), we found that the
number of code-contributing authors and unrecognized contributors has
remained relatively stable. This suggests that while the exclusion of
code contributors from authorship is not worsening, it represents a
persistent feature of scientific software development rather than a
historical artifact or transition period in research practices.
Similarly, the number of code-contributing non-authors remains constant
even as team size grows, indicating that larger research teams do not
necessarily adopt more inclusive authorship practices for code
contributors despite representing broader collaborative efforts.

\subsubsection{Modeling Article
Citations}\label{modeling-article-citations}

Building upon previous work examining the effects of team size and team
diversity on scientific impact and software quality (see
Section~\ref{sec-background}), we investigate how the number of code
contributors within a research team may be associated with an article's
research impact. We hypothesized that more code contributors might
signal greater technical complexity in research, which may be associated
with higher citation counts as the community builds upon more
technically sophisticated works (\emph{H1}). Using our filtered dataset
of article-repository pairs
(Table~\ref{tbl-team-comp-no-push-after-pub-counts}), we conducted
multiple generalized linear regression analyses to examine these
relationships while controlling for various factors. Each generalized
linear regression model included controls for the total number of
authors listed on the publication and the number of years since the date
of publication (as a decimal). Without additional controls for domain,
open access status, or article type
(Table~\ref{tbl-article-composition-overall}), our analysis revealed a
modest positive association between the number of code-contributing
authors and article citations, with each code-contributing author
associated with, on average, a $\sim$4.2\% increase in article
citations (p \textless{} 0.001).

When controlling for article type
(Table~\ref{tbl-article-composition-type}), we observed divergent
patterns between preprints and research articles. For preprints, each
code-contributing non-author was associated with a statistically
significant $\sim$3.0\% decrease in citations (p \textless{}
0.005). In contrast, research articles showed more positive
associations: we found a significant positive relationship between
code-contributing authors and citations (p \textless{} 0.001), though we
cannot estimate the precise magnitude due to the non-significant main
effect in the model. Additionally, each code-contributing non-author was
associated with a $\sim$0.1\% increase in expected citations
for research articles (p \textless{} 0.001).

Based on these findings, we \textbf{fail to reject} our hypothesis
(\emph{H1}) that ``the number of individuals contributing code to a
publication's associated repository positively correlates with the
article's citation count.'' It is important to note that, the
relationship between contributors and citations is statistically
significant, but these effects are modest in magnitude, and differ
substantially between research articles (positive association) and
preprints (negative association for non-author code contributors). These
variations suggest that the relationship between code contributions and
citation impact is context-dependent and more complex than we originally
stated in our hypothesis. We return to this finding in the discussion
and limitations sections that follow.

\subsection{Characteristics of Scientific Code
Contributors}\label{characteristics-of-scientific-code-contributors}

\subsubsection{Author Positions of Code Contributing
Authors}\label{author-positions-of-code-contributing-authors}

Building upon previous work examining the relationship between
authorship position and research contributions, we investigate how
author position may relate to code contribution patterns. We
hypothesized that first authors, traditionally contributing the bulk of
intellectual and experimental work, are most likely to contribute code
to a project (\emph{H2a}).

To analyze these patterns within our previously filtered dataset of
article-repository pairs
(Table~\ref{tbl-team-comp-no-push-after-pub-counts}), we conducted
Chi-square tests of independence between author position and code
contribution status. These tests revealed significant associations
between author position and likelihood of code contribution overall and
when controlling for research domain, article type, and open access
status (all p \textless{} 0.01), indicating that the proportion of
authors contributing code differs significantly based on author
position\footnote{Community practices and norms around authorship order
  vary across disciplines and over time. One particular changing
  practice is the adoption of shared first authorship, where two or more
  authors are designated as having contributed equally to the work.
  However, after all filtering required for analysis was complete, none
  of the remaining articles included multiple first authors.}. Following
these significant associations, we examined the specific proportions
across positions (Table~\ref{tbl-post-hoc-tests-on-author-positions}):
69.8\% of first authors contributed code to their projects, compared to
only 9.7\% of middle authors and 7.6\% of last authors. The differences
in these proportions remained statistically significant across all
tested scenarios, regardless of research domain, article type, or open
access status.

Based on these findings, we \textbf{\emph{fail to reject}} our
hypothesis (\emph{H2a}) that ``first authors have higher code
contribution rates than authors in other positions.'' The data
demonstrates that the proportion of first authors who contribute code
(69.8\%) is significantly higher than the proportion of both middle
authors (9.7\%) and last authors (7.6\%). This relationship remains
robust and statistically significant across all tested conditions,
including variations in research domain, article type, and open access
status, indicating a fundamental connection between authorship position
and technical contribution in scientific research.

\subsubsection{Corresponding Status of Code Contributing
Authors}\label{corresponding-status-of-code-contributing-authors}

Building upon our analysis of author position, we next examine how
corresponding author status relates to code contribution patterns. We
hypothesized that corresponding authors, who traditionally maintain
research artifacts and serve as primary points of contact, would be more
likely to contribute code compared to non-corresponding authors
(\emph{H2b})

To analyze these relationships within our filtered dataset of
article-repository pairs, we conducted Chi-square tests of independence
between corresponding author status and code contribution status. Our
analysis revealed patterns contrary to our initial hypothesis. The
proportion of code contributors was low among both groups, with only
29.7\% of corresponding authors and 20.6\% of non-corresponding authors
contributing code to their projects. Further examination
(Table~\ref{tbl-post-hoc-tests-on-corresponding-status}) showed that
this pattern holds across nearly all conditions, with only one
exception: corresponding authors in closed-access publications showed no
significant difference in their proportion of code contributors.
However, this was tested with a sample of less than 200 authors.

Based on these findings, we \textbf{\emph{reject}} our hypothesis
(\emph{H2b}) that ``corresponding authors have higher code contribution
rates than non-corresponding authors.'' Contrary to our expectations,
our analysis revealed that the proportion of code contributors among
corresponding authors (29.7\%) did not significantly differ from the
proportion among non-corresponding authors (20.6\%). This pattern of
similar proportions remained consistent across most studied conditions,
with a single exception in closed-access publications which we believe
is due to the relatively small sample size available for testing
(n=194).

\subsubsection{Modeling Author H-Index}\label{modeling-author-h-index}

Building upon previous work examining career implications for
researchers who prioritize software development (see
Section~\ref{sec-background}), we investigated how varying levels of
code contribution relate to scholarly impact through h-index metrics. To
ensure a robust analysis, we applied several key data filtering steps.
We only included researchers with at least three publications in our
dataset, removed those with more than three developer account
associations, and used each researcher's most common domain, article
type, and author position, with ties broken by the most recent
occurrence. We removed h-index outliers by excluding researchers below
the bottom 3rd and above the top 97th percentiles. Finally, we removed
any author-developer-account pairs with predictive model confidence of
less than 0.97. Table~\ref{tbl-h-index-counts} summarizes the number of
researchers in each coding frequency group, categorized by author
position, publication type, and research domain.

We categorized researchers' coding contributions into mutually exclusive
groups: non-coders (no code contributions), any coding (code
contribution in less than half of article-repository pairs), majority
coding (code contribution in at least half, but not all,
article-repository pairs), and always coding (code contribution in every
article-repository pair).

\begin{figure}

\centering{

\includegraphics[width=\textwidth]{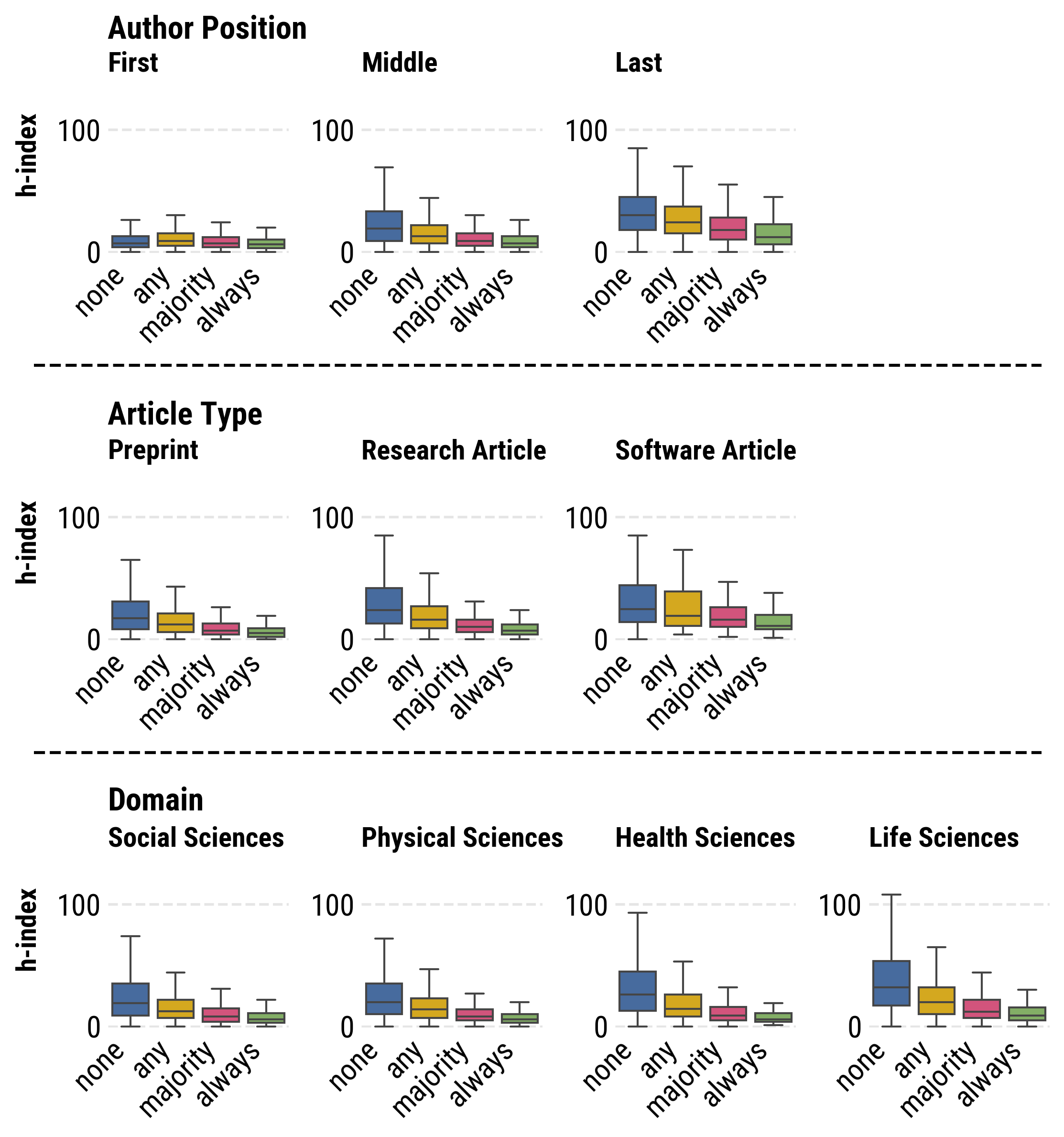}

}

\caption{\label{fig-author-h-index-by-coding-status}Distribution of
author h-index by coding frequency across three key publication factors.
Results are grouped by each author's most frequent: (1) position in
publication bylines (first, middle, or last), (2) publication type
(preprint, research article, or software article), and (3) research
domain (Social Sciences, Physical Sciences, Health Sciences, or Life
Sciences). Within each subplot, h-indices are divided by the author's
coding frequency: `none' (no coding in any of their publications), `any'
(coding in at least one but fewer than half of their publications),
`majority' (coding in at least half but not all of their publications),
and `always' (coding in each of their publications). Authors are only
included if they have three or more publications within our dataset and
are associated with no more than three developer accounts, with each
association having a predicted model confidence of at least 97\%.}

\end{figure}

Figure~\ref{fig-author-h-index-by-coding-status} shows the distribution
of author h-indices across these coding frequency groups, grouped by
author position, publication type, and research domain.

Using multiple generalized linear regressions, and while controlling for
an author's total number of published works, we find a consistent and
statistically significant negative relationship between code
contribution frequency and h-index across multiple controls. Our initial
analysis, controlled only by an author's publication count
(Table~\ref{tbl-researcher-coding-status-no-control}) indicates
increasingly adverse h-index effects as researcher coding frequency
increases. Compared to non-coding authors, researchers were associated
with progressively lower h-indices: authors who occasionally provided
code showed a $\sim$27.3\% lower h-index (p \textless{} 0.001),
authors who provided code for a majority of their works demonstrated a
$\sim$53.5\% lower h-index (p \textless{} 0.001), and always
coding authors exhibited a $\sim$62.1\% lower h-index (p
\textless{} 0.001).

When controlling for author position
(Table~\ref{tbl-researcher-coding-status-author-position}), we found a
general pattern of reduced h-indices with increased code contribution,
with one notable exception. Occasional coding first authors were
associated with a $\sim$14.9\% higher h-index (p \textless{}
0.001), while always coding first authors saw a $\sim$21.6\%
reduction compared to non-coding first authors (p \textless{} 0.001).
For middle and last authors, the pattern was more consistently negative.
Middle authors who occasionally coded showed a $\sim$26.6\%
lower h-index (p \textless{} 0.001), and those who always coded
demonstrated a $\sim$52.9\% lower h-index (p \textless{}
0.001). Similarly, last authors who occasionally coded experienced a
$\sim$13.1\% lower h-index (p \textless{} 0.001), with always
coding last authors showing a $\sim$45.7\% lower h-index (p
\textless{} 0.001).

When controlling for research domain
(Table~\ref{tbl-researcher-coding-status-domain}), majority coding
scientists showed significant h-index reductions across all domains.
Health sciences researchers saw the most dramatic reduction at
$\sim$76.5\% (p \textless{} 0.001), followed by physical
sciences at $\sim$52.6\% (p \textless{} 0.001), social sciences
at $\sim$51.4\% (p \textless{} 0.001), and life sciences at
$\sim$47.1\% (p \textless{} 0.001).

Analyzing by common article type
(Table~\ref{tbl-researcher-coding-status-article-type}) revealed similar
patterns. For authors primarily publishing preprints, the h-index
reductions were substantial: $\sim$25.6\% for occasional
coding, $\sim$53.5\% for majority coding, and
$\sim$62.9\% for always coding authors. Authors primarily
publishing software articles showed slightly better but still
significant reductions: $\sim$33.1\% for majority coding and
$\sim$33.0\% for always coding authors.

Based on these findings, we \textbf{\emph{fail to reject}} our
hypothesis (\emph{H3}) that ``the frequency with which individual
researchers contribute code to their research projects is negatively
correlated with their h-index.'' Our analysis demonstrates a clear and
statistically significant negative relationship between code
contribution frequency and scholarly impact as measured by the
researcher's h-index. This relationship was robust across multiple
analytical controls, including author position, research domain, and
article type. These results are particularly striking because our models
include publication count as an input feature, suggesting that these
h-index reductions persist even when accounting for total research
output.

\section{Discussion}\label{sec-discussion}

Our analysis reveals significant disparities in the recognition of
software contributions to scientific research, with
$\sim$28.6\% (n=5,694) of articles having code-contributors not
matched to any article author, partially representing unrecognized code
contribution. This persistent pattern over time and team size suggests a
systemic disconnect between software development and scientific
recognition systems, reflecting challenges in how scientific
contributions are valued and credited. This exclusion reflects what
\citet{shapin1989invisible} observed about scientific authority---the
selective attribution of technical work as either genuine knowledge or
mere skill significantly impacts who receives formal recognition. These
findings further support previous surveys
\citet{olivier_philippe_2019_2585783} and \citet{Carver2022ASO} where
scientists report the relegation of software contributors to either
acknowledgment sections or receiving no credit at all. The stability of
this pattern over time indicates that this phenomenon has embedded
itself in scientific software development rather than representing a
transitional phase, raising questions about scientific labor and how
reward structures integrate technical contributions.

Our finding that, on average, article-repository pairs have only a
single code contributor mirrors prior work from
\citet{farber2020analyzingGithubPapers}. Further, the distribution of
code contributions across author positions provides context to the
hierarchical organization of scientific work. First authors are
significantly more likely to contribute code with 69.8\% of all first
authors in our dataset contributing code. Middle and last authors,
meanwhile, were statistically significantly less likely to contribute
code, with only 9.7\% of middle authors and 7.6\% of last authors acting
as code-contributing members of the research team. Corresponding authors
were similarly less likely than expected to be code contributors, as we
found that within our dataset, corresponding authors were code
contributors 29.7\% of the time. These patterns align with traditional
scientific labor distribution where first authors might be expected to
handle technical aspects of research while middle and last authors are
likely specialist contributors or provide guidance and oversight
\citep{Larivire2020InvestigatingTD, sauermann2017authorship}. However,
our data did not support our initial hypothesis that corresponding
authors would also be more likely to contribute code due to their shared
responsibility for the long-term maintenance of research artifacts. This
finding suggests a potential strict division between project management
responsibilities and direct technical engagement with software
development.

The modest citation advantage associated with code-contributing authors
($\sim$4.2\% increase in citations per code-contributing
author) stands in contrast with the significant negative relationship
between coding frequency and an individual's scholarly impact (h-index).
This misalignment between code contributions and scientific recognition
creates an asymmetrical relationship in which software development may
enhance research impact but potentially penalizes individual careers.
The progressive reduction in h-index as coding frequency increases
indicates a cumulative disadvantage for frequent code contributors. This
pattern persists even when controlling for publication count, suggesting
issues in how software contributions are valued relative to other
scientific outputs. These findings echo concerns raised by
\citet{muna2016astropyproblem} about the sustainability of research
software development and highlight how current reward structures may
discourage talented developers from pursuing scientific careers.

Software development represents a form of scholarly labor that has
become increasingly essential to modern research yet remains
incompletely integrated into formal recognition systems. Similar to the
high proportion of articles with authors who made data curation
contributions towards research observed by
\citet{Larivire2020InvestigatingTD}, our finding that more than a
quarter of papers have unacknowledged code contributors highlights a
labor role that is simultaneously common and undervalued. The prevalence
of code contributions across domains demonstrates the importance of this
work to contemporary research. However, the persistent exclusion of
contributors from authorship suggests that researchers continue to
classify software development as technical support rather than
intellectual contribution. This classification may reflect disciplinary
traditions that privilege certain forms of scholarly production despite
the growing recognition that software represents a legitimate research
output \citep{Katz2020RecognizingTV}. The tension between software's
importance and contributors' recognition status raises questions about
how we define, value, and reward different forms of scientific labor in
an increasingly computational research landscape.

\subsection{Limitations}\label{limitations}

Our data collection approach introduces several methodological
constraints to consider when interpreting these results. By focusing
exclusively on GitHub repositories, we likely miss contributions stored
on alternative platforms such as GitLab, Bitbucket, or institutional
repositories, potentially skewing our understanding of contribution
patterns. As \citet{trujillo2022penumbra}, \citet{Cao2023TheRO}, and
\citet{escamilla2022riseofgithub} have all noted, while GitHub is the
predominate host of scientific software, significant portions of
research code exist on other platforms. Additionally, our reliance on
public repositories means we cannot account for private repositories or
code that were never publicly shared, potentially underrepresenting
sensitive research areas or proprietary methods.

Further, while our data processing workflow began with
$\sim$60,000 possible article-repository pairs from PLOS,
$\sim$3,000 from JOSS, $\sim$700 from SoftwareX, and
$\sim$180,000 from Papers with Code for a possible total of
$\sim$243,700\footnote{Article-repository pair counts are
  approximate because there are no snapshots of these databases at a
  single point in time. The estimates are based on counts from each data
  source taken in October 2025.}, \texttt{rs-graph-v1} contained a total
of 163,292 article-repository pairs. Many of the possible
article-repository pairs were filtered out due to the linked repository
not being accessible, or the bibliometric metadata not being available.

Our labeling of article types (software article, research article,
preprint) was based on the data source (PLOS, JOSS, SoftwareX, Papers
with Code) and in the case of Papers with Code articles, our DOI
resolution process and the document type available from OpenAlex. This
approach may misclassify certain articles, especially those from Papers
with Code (arXiv). One potential alternative approach would involve
classification of the repository itself following the recommendations of
\citet{researchsoftwareclassifications} in breaking down repositories by
their role in research (e.g., ``Modeling, Simulation, and Data
Analysis'', ``Technology Research Software'', and ``Research
Infrastructure Software''). This classification would allow us to
investigate not only the differences of ``software papers'' vs
``research articles'' and ``preprints'' (which we believe would both
typically be paired with ``Modeling, Simulation, and Data Analysis''
repositories), but the \emph{purpose} of the code as it relates to the
research. However, there is currently no established automated method
for performing this classification at scale.

Similarly, our simplification of author positions, domains, and article
types to each author's ``most common'' (most frequent, with ties broken
by most recent) introduces potential biases \citep{li2025determination}.
This reduction may obscure the diversity of an author's contributions
across different contexts, particularly for interdisciplinary
researchers or those with varied roles in different projects. Further,
these labels are created from metadata for articles only within our
dataset. That is, even though a researcher may have dozens of articles,
their ``most common'' author position, domain, and article type was
determined with data for article-repository pairs. This inherently
biases the dataset towards research teams who, as a collective,
frequently create and share software and code as a part of their
research process. While this approach was necessary for managing the
complexity of our analysis, it may not fully capture the nuances of
individual research careers.

Our predictive modeling approach for matching authors with developer
accounts presents additional limitations. The model's performance can be
affected by shorter names where less textual information is available
for matching, potentially creating biases against researchers from
cultures with shorter naming conventions. Organization accounts used for
project management pose particular challenges for accurate matching, and
while we implemented filtering mechanisms to minimize their impact, some
misclassifications may persist. Furthermore, our approach may not
capture all code contributors if multiple individuals developed code but
only one uploaded it to a repository---creating attribution artifacts
that may systematically underrepresent specific contributors,
particularly junior researchers or staff who may not have direct
repository access. However, as discussed further in the Appendix
(Section~\ref{sec-appendix-dataset-char}), our dataset is relatively
diverse, with the median preprint-repository pair having a commit
duration (the number of days between the repository's creation and the
repository's most recent commit) of 53 days, research article-repository
pairs having a median commit duration of 114 days, and software
article-repository pairs having a median commit duration of 282 days.
This diversity in commit durations suggests that our dataset contains a
range of development practices, including some ``code dumps,'' as well
as year, and multi-year long projects.

Our analytical approach required substantial data filtering to ensure
reliable results, introducing potential selection biases in our sample.
By focusing on article-repository pairs with commit activity no later
than 90 days past the date of article publication and at least three
authors and less than 11 authors, we may have systematically excluded
certain types of research projects, particularly those with extended
development timelines or extensive collaborations. Our categorization of
coding status (non-coder, any coding, majority coding, always coding)
necessarily simplifies complex contribution patterns. It does not
account for code contributions' quality, complexity, or significance.
Additionally, our reliance on OpenAlex metadata introduces certain
limitations to our analysis. While OpenAlex provides good overall
coverage, it lags behind proprietary databases in indexing references
and citations. The lag in OpenAlex data may affect our citation-based
analyses and the completeness of author metadata used in our study
\citep{alperin2024analysis}.

\subsection{Future Work}\label{future-work}

Future technical improvements may enhance our understanding of the
relationship between software development and scientific recognition
systems. Expanding analysis beyond GitHub to include other code hosting
platforms would provide a more comprehensive understanding of scientific
software development practices across domains and institutional
contexts. More sophisticated entity-matching techniques could improve
author-developer account identification, particularly for cases with
limited information or common names. Developing more nuanced measures
and classifications of code contribution type, quality, and significance
beyond binary contribution identification would better capture the true
impact of technical contributions to research (as we have started to do
in Section~\ref{sec-appendix-model-training-eval}). These methodological
advances would enable more precise tracking of how code contributions
translate---or fail to translate---into formal scientific recognition,
providing clearer evidence for policy interventions.

Our findings point to several directions for future research on the
changing nature of scientific labor and recognition. Longitudinal
studies tracking how code contribution patterns affect career
trajectories would provide valuable insights into the long-term impacts
of the observed h-index disparities and whether these effects vary
across career stages. Comparative analyses across different scientific
domains could reveal discipline-specific norms and practices around
software recognition, potentially identifying models that more equitably
credit technical contributions. Qualitative studies examining how
research teams make authorship decisions regarding code contributors
would complement our quantitative findings by illuminating the social
and organizational factors influencing recognition practices.
Additionally, to better understand corresponding authors' role in
maintaining research artifacts, future work could remove the 90-day
post-publication commit activity filter to examine long-term
sustainability actions. However, this approach must address introducing
contributors unrelated to the original paper.

Despite their growing importance, the persistent under-recognition of
software contributions suggests a need for structural interventions in
how we conceptualize and reward scientific work. Building upon efforts
like CRediT \citep{brand2015beyond}, future work should investigate
potential policy changes to better align institutional incentives with
the diverse spectrum of contributions that drive modern scientific
progress. However, as the example of CRediT demonstrates, even
well-intentioned taxonomies may reproduce existing hierarchies or create
new forms of inequality if they fail to address underlying power
dynamics in scientific communities. The challenge is not merely
technical but social: creating recognition systems that simultaneously
support innovation, ensure appropriate credit, maintain research
integrity, and foster equitable participation in an increasingly
computational scientific enterprise.

\section{Acknowledgments}\label{acknowledgments}

We thank our anonymous reviewers for their helpful feedback. We
especially thank Molly Blank, Shahan Ali Memnon, David Farr, and Jevin
West for their valuable insights at multiple stages of this research.

\section{Author Contributions}\label{author-contributions}

Eva Maxfield Brown: Conceptualization, Data Curation, Formal Analysis,
Investigation, Methodology, Software, Validation, Visualization, Writing
- original draft, Writing - review \& editing.

Isaac Slaughter: Formal Analysis, Visualization, Writing - review \&
editing.

Nicholas Weber: Conceptualization, Data Curation, Formal Analysis,
Funding Acquisition, Investigation, Methodology, Project Administration,
Resources, Supervision, Writing - original draft, Writing - review \&
editing.

\section{Competing Interests}\label{competing-interests}

The authors have no competing interests.

\section{Funding Information}\label{funding-information}

This research was supported by grants from the National Science
Foundation's (NSF) Office of Advanced Cyberinfrastructure (OAC-2211275)
and the Sloan Foundation (G-2022-19347).

\section{Data Availability}\label{data-availability}

The software and code used to train, evaluate, and apply the
author-developer account matching model is available at:
\url{https://github.com/evamaxfield/sci-soft-models}
(\url{https://doi.org/10.5281/zenodo.17401863}). The software and code
used to gather, process, and analyze the dataset of linked scientific
articles and code repositories and their associated contributors and
metadata is available at: \url{https://github.com/evamaxfield/rs-graph}
(\url{https://doi.org/10.5281/zenodo.17401960}). The compiled
rs-graph-v1 dataset, as well as data required for model training and
evaluation, is available at: \url{https://doi.org/10.7910/DVN/KPYVI1}.
Certain portions of the compiled rs-graph-v1 dataset, and the model
training and evaluation datasets contain linked personal data (e.g.,
author names, developer account usernames) and are only available by
request.

\section{References}\label{references}

\renewcommand{\bibsection}{}
\bibliography{main.bib}

\section{Appendix}\label{appendix}

\subsection{Extended Data and Methods}\label{extended-data-and-methods}

\subsubsection{Building a Dataset of Linked Scientific Articles and Code
Repositories}\label{building-a-dataset-of-linked-scientific-articles-and-code-repositories}

The increasing emphasis on research transparency has led many journals
and platforms to require or recommend code and data sharing
\citep{stodden2013toward, sharma2024analytical}, creating traceable
links between publications and code. These explicit links enable
systematic study of both article-repository and author-developer account
relationships
\citep{Hata2021ScienceSoftwareLT, Kelley2021AFF, Stankovski2024RepoFromPaperAA, milewicz2019characterizing}.

Our dataset collection process leveraged four sources of linked
scientific articles and code repositories, each with specific mechanisms
for establishing these connections:

\begin{enumerate}
\def\labelenumi{\arabic{enumi}.}
\tightlist
\item
  \textbf{Public Library of Science (PLOS)}: We extracted repository
  links from PLOS articles' mandatory data and code availability
  statements.
\item
  \textbf{Journal of Open Source Software (JOSS)}: JOSS requires
  explicit code repository submission and review as a core part of its
  publication process.
\item
  \textbf{SoftwareX}: Similar to JOSS, SoftwareX mandates code
  repositories as a publication requirement.
\item
  \textbf{Papers with Code}: This platform directly connects machine
  learning preprints with their implementations. We focus solely on the
  ``official'' article-repository relationships rather than the
  ``unverified'' or ``unofficial'' links.
\end{enumerate}

We enriched these article-repository pairs with metadata from multiple
sources to create a comprehensive and analyzable dataset. We utilized
the Semantic Scholar API for DOI resolution to ensure we found the
latest version of each article. This resolution step was particularly
important when working with preprints, as journals may have published
these papers since their inclusion in the Papers with Code dataset.
Using Semantic Scholar, we successfully resolved $\sim$56.3\%
(n=78,021) of all DOIs within our dataset\footnote{Broken out by dataset
  source, we resolved $\sim$2.1\% (n=125) of all PLOS DOIs,
  $\sim$4.0\% (n=93) of all JOSS DOIs, $\sim$0.0\%
  (n=0) of all SoftwareX DOIs, and $\sim$49.2\% (n=63,817) of
  all Papers with Code (arXiv) DOIs.}.

We then utilized the OpenAlex API to gather detailed publication
metadata, including:

\begin{itemize}
\tightlist
\item
  Publication characteristics (open access status, domain, publication
  date)
\item
  Author details (name, author position, corresponding author status)
\item
  Article- and individual-level metrics (citation counts, FWCI, h-index)
\end{itemize}

Similarly, the GitHub API provided comprehensive information for source
code repositories:

\begin{itemize}
\tightlist
\item
  Repository metadata (name, description, programming languages,
  creation date)
\item
  Contributor details (username, display name, email)
\item
  Repository-level metrics (star count, fork count, issue count)
\end{itemize}

\subsubsection{Developing a Predictive Model for Author-Developer
Account
Matching}\label{developing-a-predictive-model-for-author-developer-account-matching}

\paragraph{Annotated Dataset Creation}\label{annotated-dataset-creation}

Creating an accurate author-developer account matching model required
high quality, labeled training data that reflects real-world
identity-matching challenges. Exact matching on names or emails proved
insufficient due to variations in formatting (e.g., ``J. Doe''
vs. ``Jane Doe''), use of institutional versus personal email addresses,
and incomplete information. However, author and developer account
information often contains sufficient similarities for probabilistic
matching, such as when author ``Jane Doe'' corresponds to username
``jdoe'' or ``janedoe123.''

To efficiently build our training and evaluation dataset, we used JOSS
articles as we believed they typically feature higher
author-developer-account overlap, increasing positive match density. Our
dataset creation process followed these steps:

\begin{enumerate}
\def\labelenumi{\arabic{enumi}.}
\tightlist
\item
  We generated semantic embeddings for each developer account and author
  name using the
  \href{https://huggingface.co/sentence-transformers/multi-qa-MiniLM-L6-cos-v1}{multi-qa-MiniLM-L6-cos-v1}
  model from the Sentence Transformers Python library
  \citep{reimers-2019-sentence-bert}.
\item
  We calculated cosine similarity between all potential
  author-developer-account pairs for each article-repository pair.
\item
  We selected the three most similar authors for each developer account
  for annotation efficiency.
\end{enumerate}

From these generated author-developer-account pairs, we randomly
selected 2,999 for classification by two independent annotators as
either matches or non-matches, resolving disagreements through
discussion and verification. The resulting dataset contains 451
($\sim$15.0\%) positive matches and 2,548
($\sim$85.0\%) negative matches, comprising 2,027 unique
authors and 2,733 unique developer accounts.

Our collected data for annotation confirmed that exact matching would be
insufficient---only 2,191 ($\sim$80.2\%) of developer accounts
had associated display names and just 839 ($\sim$30.7\%) had
associated email addresses.

\paragraph{Training and
Evaluation}\label{sec-appendix-model-training-eval}

Our training and evaluation methodology began with careful dataset
preparation to prevent data leakage between training and test sets. To
ensure complete separation of authors and developers, we randomly
selected 10\% of unique authors and 10\% of unique developers,
designating any pairs containing these selected entities for the test
set. This entity-based splitting strategy resulted in 2,442
($\sim$81.4\%) pairs for training and 557
($\sim$18.6\%) pairs for testing.

For our predictive model, we evaluated three transformer-based
architectures that have demonstrated strong performance in
entity-matching tasks:

\begin{itemize}
\tightlist
\item
  DeBERTa-v3-base \citep{he2021debertav3, he2021deberta}
\item
  mBERT (bert-base-multilingual-cased) \citep{bert2018}
\item
  DistilBERT \citep{Sanh2019DistilBERTAD}
\end{itemize}

We systematically evaluated these base models across different
combinations of developer-account features, ranging from using only the
username to incorporating complete profile information (username,
display name, and email address). We fine-tuned all models using the
Adam optimizer with a linear learning rate of 1e-5 for training and a
batch size of 8 for training and evaluation. Given the size of our
dataset and the binary nature of our classification task, models were
trained for a single epoch to prevent overfitting.

We evaluated model performance using precision, recall, and F1-score.
This evaluation framework allowed us to directly compare model
architectures and feature combinations while accounting for the balance
between precision and recall in identifying correct matches.

Our comprehensive model evaluation revealed that fine-tuning
DeBERTa-v3-base \citep{he2021debertav3} with developer username and
display name as input features produces optimal performance for
author-developer matching. This model configuration achieved a binary F1
score of 0.944, with an accuracy of 0.984, precision of 0.938, and
recall of 0.95. Table~\ref{tbl-em-model-comparison} presents a complete
comparison of model architectures and feature combinations.

\begin{table}

\caption{\label{tbl-em-model-comparison}Comparison of Models for Author-Developer-Account Matching}

\centering{

\centering

\begin{tabular}{llrrrr}
\toprule
\textbf{Optional Feats.} & \textbf{Model} & \textbf{Accuracy} & \textbf{Precision} & \textbf{Recall} & \textbf{F1} \\
\midrule
    \cellcolor{white}name & \cellcolor{white}deberta & \cellcolor{white}0.984 & \cellcolor{white}0.938 & \cellcolor{white}0.950 & \cellcolor{white}0.944 \\
    \cellcolor{gray!10}name, email & \cellcolor{gray!10}bert-multilingual & \cellcolor{gray!10}0.984 & \cellcolor{gray!10}0.938 & \cellcolor{gray!10}0.950 & \cellcolor{gray!10}0.944 \\
    \cellcolor{white}name, email & \cellcolor{white}deberta & \cellcolor{white}0.982 & \cellcolor{white}0.907 & \cellcolor{white}0.975 & \cellcolor{white}0.940 \\
    \cellcolor{gray!10}name & \cellcolor{gray!10}bert-multilingual & \cellcolor{gray!10}0.982 & \cellcolor{gray!10}0.938 & \cellcolor{gray!10}0.938 & \cellcolor{gray!10}0.938 \\
    \cellcolor{white}name & \cellcolor{white}distilbert & \cellcolor{white}0.978 & \cellcolor{white}0.936 & \cellcolor{white}0.912 & \cellcolor{white}0.924 \\
    \cellcolor{gray!10}name, email & \cellcolor{gray!10}distilbert & \cellcolor{gray!10}0.978 & \cellcolor{gray!10}0.936 & \cellcolor{gray!10}0.912 & \cellcolor{gray!10}0.924 \\
    \cellcolor{white}email & \cellcolor{white}deberta & \cellcolor{white}0.957 & \cellcolor{white}0.859 & \cellcolor{white}0.838 & \cellcolor{white}0.848 \\
    \cellcolor{gray!10}email & \cellcolor{gray!10}bert-multilingual & \cellcolor{gray!10}0.950 & \cellcolor{gray!10}0.894 & \cellcolor{gray!10}0.738 & \cellcolor{gray!10}0.808 \\
    \cellcolor{white}n/a & \cellcolor{white}deberta & \cellcolor{white}0.946 & \cellcolor{white}0.847 & \cellcolor{white}0.762 & \cellcolor{white}0.803 \\
    \cellcolor{gray!10}n/a & \cellcolor{gray!10}bert-multilingual & \cellcolor{gray!10}0.941 & \cellcolor{gray!10}0.862 & \cellcolor{gray!10}0.700 & \cellcolor{gray!10}0.772 \\
    \cellcolor{white}n/a & \cellcolor{white}distilbert & \cellcolor{white}0.856 & \cellcolor{white}0.000 & \cellcolor{white}0.000 & \cellcolor{white}0.000 \\
    \cellcolor{gray!10}email & \cellcolor{gray!10}distilbert & \cellcolor{gray!10}0.856 & \cellcolor{gray!10}0.000 & \cellcolor{gray!10}0.000 & \cellcolor{gray!10}0.000 \\
\bottomrule
\end{tabular}

}

\end{table}%

Analysis of each model's performance revealed that including developer
display names had the most significant positive impact on model
performance compared to username alone. We also observed that mBERT's
performance was comparable to DeBERTa's while using the developer email
address as an additional input feature. However, we selected the DeBERTa
configuration as it consistently performed well across various feature
combinations.

To facilitate the reuse of our work, we have made our trained model and
supporting code publicly available. Complete fine-tuning, evaluation,
and inference code is available as the Python package:
\href{https://github.com/evamaxfield/sci-soft-models}{sci-soft-models},
and the fine-tuned model has been released on HuggingFace
(\href{https://huggingface.co/evamxb/dev-author-em-clf}{evamxb/dev-author-em-clf}).

\paragraph{Evaluation of Model Performance on Non-JOSS Author-Developer
Pairs}\label{evaluation-of-model-performance-on-non-joss-author-developer-pairs}

To assess our model's generalizability beyond the JOSS dataset used for
training, we conducted an additional evaluation using
author-developer-account pairs from the PLOS, SoftwareX, and Papers with
Code datasets. We created a dataset of all combinations of possible
author-developer pairs from 20 article-repository pairs from each
dataset. This resulted in 535 possible author-developer pairs for
annotation across the 60 total article-repository pairs. Two independent
annotators classified each of the possible author-developer pairs as
either matches or non-matches. This annotation process mirrors how our
complete dataset was constructed by using the trained model to predict
matches from all possible author-developer pairs for a given
article-repository pair. The two annotators achieved a Cohen's kappa of
0.948, or ``almost perfect'' agreement
\citep{cohen1960coefficient, mchugh2012interrater}. Annotators further
discussed and resolved all disagreements.

The trained author-developer matching model was then applied to this new
dataset, achieving an overall binary-F1 score of 0.89, precision of 0.92
and recall of 0.87 (positive=``match''). The overall macro-F1 score was
0.94, with a precision of 0.95 and recall of 0.93. Per-dataset
performance is presented in Table~\ref{tbl-em-model-non-joss-eval}.
Across each of the non-JOSS datasets, the model demonstrated strong
performance, with binary-F1 scores ranging from 0.88 to 0.90, and
acheiving 0.94 for all macro-F1 scores.

\begin{longtable}[]{@{}
  >{\raggedright\arraybackslash}p{(\linewidth - 12\tabcolsep) * \real{0.1748}}
  >{\raggedright\arraybackslash}p{(\linewidth - 12\tabcolsep) * \real{0.1068}}
  >{\raggedright\arraybackslash}p{(\linewidth - 12\tabcolsep) * \real{0.1748}}
  >{\raggedright\arraybackslash}p{(\linewidth - 12\tabcolsep) * \real{0.1456}}
  >{\raggedright\arraybackslash}p{(\linewidth - 12\tabcolsep) * \real{0.0971}}
  >{\raggedright\arraybackslash}p{(\linewidth - 12\tabcolsep) * \real{0.1650}}
  >{\raggedright\arraybackslash}p{(\linewidth - 12\tabcolsep) * \real{0.1359}}@{}}
\caption{Per-dataset performance of the author-developer-account
matching model on non-JOSS
datasets.}\label{tbl-em-model-non-joss-eval}\tabularnewline
\toprule\noalign{}
\begin{minipage}[b]{\linewidth}\raggedright
dataset
\end{minipage} & \begin{minipage}[b]{\linewidth}\raggedright
binary-F1
\end{minipage} & \begin{minipage}[b]{\linewidth}\raggedright
binary-Precision
\end{minipage} & \begin{minipage}[b]{\linewidth}\raggedright
binary-Recall
\end{minipage} & \begin{minipage}[b]{\linewidth}\raggedright
macro-F1
\end{minipage} & \begin{minipage}[b]{\linewidth}\raggedright
macro-Precision
\end{minipage} & \begin{minipage}[b]{\linewidth}\raggedright
macro-Recall
\end{minipage} \\
\midrule\noalign{}
\endfirsthead
\toprule\noalign{}
\begin{minipage}[b]{\linewidth}\raggedright
dataset
\end{minipage} & \begin{minipage}[b]{\linewidth}\raggedright
binary-F1
\end{minipage} & \begin{minipage}[b]{\linewidth}\raggedright
binary-Precision
\end{minipage} & \begin{minipage}[b]{\linewidth}\raggedright
binary-Recall
\end{minipage} & \begin{minipage}[b]{\linewidth}\raggedright
macro-F1
\end{minipage} & \begin{minipage}[b]{\linewidth}\raggedright
macro-Precision
\end{minipage} & \begin{minipage}[b]{\linewidth}\raggedright
macro-Recall
\end{minipage} \\
\midrule\noalign{}
\endhead
\bottomrule\noalign{}
\endlastfoot
PLOS & 0.88 & 0.79 & 1.00 & 0.94 & 0.89 & 0.99 \\
SoftwareX & 0.90 & 0.96 & 0.85 & 0.94 & 0.97 & 0.92 \\
Papers With Code & 0.89 & 1.00 & 0.81 & 0.94 & 0.98 & 0.90 \\
\end{longtable}

\paragraph{Model Limitations}\label{model-limitations}

While our model demonstrates strong performance, we acknowledge certain
limitations in our approach:

\begin{enumerate}
\def\labelenumi{\arabic{enumi}.}
\tightlist
\item
  \textbf{Short name sensitivity}: Shorter names (both usernames and
  display names) can affect the model's performance, as less textual
  information is available for matching.
\item
  \textbf{Organization accounts}: Research lab accounts used for project
  management present a potential challenge for accurate matching, as
  they do not correspond to individual authors. However, our filtering
  mechanisms applied before analysis help minimize their impact on
  modeling.
\end{enumerate}

Additional limitations are discussed in a qualitative error analysis
conducted as part of our evaluation of the code-contributing non-author
sub-sample in Section~\ref{sec-appendix-cc-na-error-analysis}.

\subsubsection{Dataset Characteristics and Repository
Types}\label{sec-appendix-dataset-char}

Our compiled dataset appears to contain a mix of repository types,
varying from analysis script repositories to software tools and likely
some ``code dumps'' (where code is copied to a new repository
immediately before publication). This diversity is reflected in the
commit duration patterns across different publication types. The median
commit duration for repositories in our analysis is:

\begin{itemize}
\tightlist
\item
  53 days for preprints
\item
  114 days for research articles
\item
  282 days for software articles
\end{itemize}

Complete statistics on commit durations, including count, mean, and
quantile details, are available in
Table~\ref{tbl-commit-duration-distributions}.

\begin{longtable}[]{@{}lllllllllll@{}}

\caption{\label{tbl-commit-duration-distributions}Commit duration (in
days) distributions for different publication types. Only includes
article-repository pairs with a most recent commit no later than 90 days
after publication and excludes publications from research teams in the
top 3\% of total author sizes.}

\tabularnewline

\toprule\noalign{}
& count & mean & std & min & 10\% & 25\% & 50\% & 75\% & 90\% & max \\
article\_type & & & & & & & & & & \\
\midrule\noalign{}
\endhead
\bottomrule\noalign{}
\endlastfoot
preprint & 2683 & 110 & 182 & -1520 & 0 & 6 & 53 & 138 & 285 & 2091 \\
research article & 17017 & 193 & 253 & -931 & 0 & 19 & 114 & 269 & 487 &
3176 \\
software article & 200 & 394 & 475 & -1 & 0 & 50 & 282 & 536 & 951 &
3007 \\

\end{longtable}

\subsection{Distributions of Author-Developer-Account Prediction
Confidence}\label{distributions-of-author-developer-account-prediction-confidence}

Thresholding the predictive model confidence at 0.97 resulted in a
$\sim$3.2\% (n=2,909) reduction in the number of
author-developer-account pairs (from an unfiltered total of 90,086
author-developer-account pairs). This threshold was chosen to ensure a
high level of confidence in the matches while retaining a large number
of pairs for analysis.

\subsection{Filtered Dataset Description for Article-Citation,
Author-Position, and Author-Correspondence
Analysis}\label{filtered-dataset-description-for-article-citation-author-position-and-author-correspondence-analysis}

\begin{table}

\caption{\label{tbl-team-comp-no-push-after-pub-counts}Counts of article-repository pairs, authors, and developers for research teams. Only includes research teams from article-repository pairs with a most recent commit no later than 90 days after publication and excludes research teams in the top 3\% of total author sizes.}

\centering{

    \centering
    \small

    \begin{tabular}{llrrr}
    \toprule
    \textbf{Category} & \textbf{Subset} & \textbf{Article-Repository Pairs} & \textbf{Authors} & \textbf{Developers} \\
    \midrule
        \multirow{4}{*}{\textbf{By Domain}} & \cellcolor{gray!10}Health Sciences & \cellcolor{gray!10}929 & \cellcolor{gray!10}4,904 & \cellcolor{gray!10}1,181 \\
    & Life Sciences & 1,685 & 7,910 & 2,257 \\
    & \cellcolor{gray!10}Physical Sciences & \cellcolor{gray!10}16,164 & \cellcolor{gray!10}55,161 & \cellcolor{gray!10}20,904 \\
    & Social Sciences & 1,122 & 4,661 & 1,562 \\\midrule
    \multirow{3}{*}{\textbf{By Document Type}} & \cellcolor{gray!10}preprint & \cellcolor{gray!10}2,683 & \cellcolor{gray!10}11,327 & \cellcolor{gray!10}4,190 \\
    & research article & 17,017 & 61,279 & 21,482 \\
    & \cellcolor{gray!10}software article & \cellcolor{gray!10}200 & \cellcolor{gray!10}905 & \cellcolor{gray!10}449 \\\midrule
    \multirow{2}{*}{\textbf{By Access Status}} & \cellcolor{gray!10}Closed & \cellcolor{gray!10}1,070 & \cellcolor{gray!10}5,054 & \cellcolor{gray!10}1,679 \\
    & Open & 18,830 & 66,677 & 24,026 \\\midrule
    \multirow{4}{*}{\textbf{By Data Source}} & \cellcolor{gray!10}joss & \cellcolor{gray!10}77 & \cellcolor{gray!10}353 & \cellcolor{gray!10}246 \\
    & plos & 2,109 & 10,345 & 2,729 \\
    & \cellcolor{gray!10}pwc & \cellcolor{gray!10}17,591 & \cellcolor{gray!10}59,261 & \cellcolor{gray!10}22,313 \\
    & softwarex & 123 & 554 & 204 \\\midrule
    \textbf{Total} & & \textbf{19,900} & \textbf{69,808} & \textbf{25,358} \\
    \bottomrule
    \end{tabular}

}

\end{table}%

\subsubsection{Distributions of Team
Composition}\label{distributions-of-team-composition}

Table~\ref{tbl-team-composition-counts} provides detailed statistics on
the composition of research teams across different domains, article
types, and open access statuses.

\begin{table}

\caption{\label{tbl-team-composition-counts}Mean and Standard Deviation of Non-Code-Contributing Authors (NCC-A), Code-Contributing Authors (CC-A), and Code-Contributing Non-Authors (CC-NA) Research Team Members by Domain, Article Type, and Open Access Status. Only includes research teams from article-repository pairs with repositories that have programming language files, a most recent commit no later than 90 days after publication, and excludes research teams in the top $3\%$ of total author sizes.}

\centering{

\centering
\small

\begin{tabular}{llrrrr}
\toprule\textbf{Control} & \textbf{Subset} & \textbf{Total Authors} & \textbf{NCC-A} & \textbf{CC-A} & \textbf{CC-NA} \\
\midrule
    \multirow{2}{*}{\textbf{OA Status}} & \cellcolor{gray!10}Closed & \cellcolor{gray!10}5.1 $\pm$ 1.8 & \cellcolor{gray!10}4.0 $\pm$ 1.9 & \cellcolor{gray!10}1.1 $\pm$ 0.7 & \cellcolor{gray!10}0.6 $\pm$ 2.1 \\
    & Open & 4.8 $\pm$ 1.8 & 3.8 $\pm$ 1.9 & 1.0 $\pm$ 0.7 & 0.5 $\pm$ 1.8 \\\midrule
    \multirow{4}{*}{\textbf{Domain}} & \cellcolor{gray!10}Health Sciences & \cellcolor{gray!10}5.8 $\pm$ 2.2 & \cellcolor{gray!10}4.8 $\pm$ 2.3 & \cellcolor{gray!10}1.0 $\pm$ 0.6 & \cellcolor{gray!10}0.4 $\pm$ 1.3 \\
    & Life Sciences & 5.0 $\pm$ 2.0 & 4.0 $\pm$ 2.1 & 1.0 $\pm$ 0.7 & 0.4 $\pm$ 1.2 \\
    & \cellcolor{gray!10}Physical Sciences & \cellcolor{gray!10}4.8 $\pm$ 1.7 & \cellcolor{gray!10}3.7 $\pm$ 1.8 & \cellcolor{gray!10}1.0 $\pm$ 0.7 & \cellcolor{gray!10}0.5 $\pm$ 1.9 \\
    & Social Sciences & 4.5 $\pm$ 1.7 & 3.4 $\pm$ 1.7 & 1.1 $\pm$ 0.7 & 0.3 $\pm$ 1.2 \\\midrule
    \multirow{3}{*}{\textbf{Article Type}} & \cellcolor{gray!10}preprint & \cellcolor{gray!10}4.8 $\pm$ 1.7 & \cellcolor{gray!10}3.7 $\pm$ 1.8 & \cellcolor{gray!10}1.0 $\pm$ 0.7 & \cellcolor{gray!10}0.6 $\pm$ 2.3 \\
    & research article & 4.8 $\pm$ 1.8 & 3.8 $\pm$ 1.9 & 1.0 $\pm$ 0.7 & 0.4 $\pm$ 1.7 \\
    & \cellcolor{gray!10}software article & \cellcolor{gray!10}4.6 $\pm$ 1.7 & \cellcolor{gray!10}3.1 $\pm$ 1.8 & \cellcolor{gray!10}1.5 $\pm$ 1.3 & \cellcolor{gray!10}0.9 $\pm$ 1.1 \\\midrule
    \multirow{1}{*}{\textbf{Overall}} & \textbf{} & \textbf{4.8 $\pm$ 1.8} & \textbf{3.8 $\pm$ 1.9} & \textbf{1.0 $\pm$ 0.7} & \textbf{0.5 $\pm$ 1.8} \\
\bottomrule
\end{tabular}

}

\end{table}%

\subsection{Additional Analysis of Code-Contributing
Non-Authors}\label{sec-appendix-additional-cc-na-analysis}

\subsubsection{Sample Construction and
Labeling}\label{sample-construction-and-labeling}

To better understand the nature and extent of contributions made by
code-contributing non-authors in our dataset, we conducted a detailed
analysis using a random sample of 200 individuals from our filtered
dataset used in other analyses
(Table~\ref{tbl-team-comp-no-push-after-pub-counts}). Our analysis
combined qualitative labeling of contribution types with quantitative
analysis of commit activity, additions, and deletions made by these
contributors to their respective article-repository pairs.

\paragraph{Annotation Process}\label{annotation-process}

Two independent annotators labeled each of the 200 code-contributing
non-authors across three dimensions after completing two rounds of trial
labeling on 20 cases to establish agreement. The final labeling criteria
were:

\textbf{Contribution Type}:

\begin{itemize}
\tightlist
\item
  \emph{``docs''}: Contributors who only modified documentation files
  (README, LICENSE, etc.) or made changes limited to code comments.
\item
  \emph{``code''}: Contributors who modified actual code files (.py, .R,
  .js, etc.) with substantive changes or modified code support files
  (requirements.txt, pyproject.toml, package.json, etc.). Contributors
  who made code and documentation changes were labeled ``code.''
\item
  \emph{``other''}: Contributors whose changes did not fit the above
  categories, including those who committed to upstream forks or merged
  code without authoring it.
\end{itemize}

\textbf{Author Matching Assessment}:

\begin{itemize}
\tightlist
\item
  \emph{``yes''}: Contributors who should have been matched to an author
  (missed classification).
\item
  \emph{``no''}: Contributors correctly classified as non-authors.
\item
  \emph{``unclear''}: Cases with insufficient information for
  determination.
\end{itemize}

\textbf{Bot Account Detection}:

\begin{itemize}
\tightlist
\item
  \emph{``yes''}: Automated accounts (GitHub Actions, Dependabot, etc.).
\item
  \emph{``no''}: Human users.
\end{itemize}

After establishing near perfect agreement for contribution type
($\kappa=0.89$), and perfect agreement for author matching assessment and bot
account detection ($\kappa=1.0$), each annotator independently labeled 90
contributors--- the final sample of 200 created by combining both sets
plus the 20 cases used for criteria development.

\paragraph{Quantitative Metrics}\label{quantitative-metrics}

For each code-contributing non-author, we collected commit activity data
using the GitHub API contributor stats endpoint:

\begin{itemize}
\tightlist
\item
  \textbf{Number of Commits}: The total number of commits made by the
  code contributor to the article-repository pair.
\item
  \textbf{Number of Additions}: The total number of lines of code added
  by the code contributor to the article-repository pair.
\item
  \textbf{Number of Deletions}: The total number of lines of code
  deleted by the code contributor to the article-repository pair.
\item
  \textbf{Number of Total Repository Commits}: The total number of
  commits made to the article-repository pair, regardless of code
  contributor.
\item
  \textbf{Number of Total Repository Additions}: The number of lines of
  code added to the article-repository pair, regardless of code
  contributor.
\item
  \textbf{Number of Total Repository Deletions}: The total number of
  lines of code deleted from the article-repository pair, regardless of
  code contributor.
\end{itemize}

We additionally calculated the absolute change for each code contributor
as the sum of additions and deletions, which provides a measure of the
total impact of their contributions. Further, we normalized these
metrics by the total number of commits, additions, deletions, and
absolute changes made to the article-repository pair, regardless of code
contributor. This normalization allows us to compare the relative
contribution of each code-contributing non-author to the overall amount
of changes to the repository.

\subsubsection{Results}\label{results}

We find that $\sim$39\% (n=78) of code-contributing non-authors
were correctly classified as non-authors, $\sim$30.5\% (n=61)
were unclear due to insufficient profile information, and
$\sim$30.5\% (n=61) appeared to be missed classifications that
should have been matched to authors. Only two accounts
($\sim$1\%) were identified as bot accounts.

When broken out by contribution type, we find that:

\begin{itemize}
\tightlist
\item
  \emph{``true non-authors'' (n=78)}: 59 contributed code, 13
  contributed documentation, and 4 contributed some other type of change
\item
  \emph{``missed classifications'' (n=61)}: 49 contributed code, 12
  contributed documentation, and 0 contributed some other type of change
\item
  \emph{``unclear'' (n=61)}: 50 contributed code, 8 contributed
  documentation, and 3 contributed some other type of change
\end{itemize}

Table~\ref{tbl-true-cc-na-commit-stats} and
Table~\ref{tbl-unclear-cc-na-commit-stats} present commit statistics for
true non-authors and unclear cases, respectively. Among true non-authors
making code contributions, the top quartile (75th percentile and above)
contributed $\sim$10.7\% of total repository commits and
$\sim$14.4\% of absolute changes (additions + deletions). The
unclear cases showed substantially higher contribution levels. Code
contributors in this group comprised $\sim$50.5\% of total
repository commits and $\sim$41.7\% of repository absolute
changes, even at the 25th percentile.

\begin{longtable}[]{@{}llllllll@{}}

\caption{\label{tbl-true-cc-na-commit-stats}Commit statistics for
code-contributing non-authors labeled as true non-authors. Statistics
are calculated as the proportion of commits, additions, deletions, and
absolute changes made by the code-contributing non-author to the total
commits, additions, deletions, and absolute changes made to the
article-repository pair, regardless of code-contributor.}

\tabularnewline

\toprule\noalign{}
& mean & std & min & 25\% & 50\% & 75\% & max \\
\midrule\noalign{}
\endhead
\bottomrule\noalign{}
\endlastfoot
commit\_stats & 0.178 & 0.307 & 0.001 & 0.007 & 0.029 & 0.107 & 1.0 \\
addition\_stats & 0.227 & 0.380 & 0.000 & 0.001 & 0.007 & 0.192 & 1.0 \\
deletion\_stats & 0.193 & 0.358 & 0.000 & 0.000 & 0.006 & 0.149 & 1.0 \\
abs\_stats & 0.222 & 0.379 & 0.000 & 0.001 & 0.010 & 0.144 & 1.0 \\

\end{longtable}

\begin{longtable}[]{@{}llllllll@{}}

\caption{\label{tbl-unclear-cc-na-commit-stats}Commit statistics for
code-contributing non-authors labeled as unclear. Statistics are
calculated as the proportion of commits, additions, deletions, and
absolute changes made by the code-contributing non-author to the total
commits, additions, deletions, and absolute changes made to the
article-repository pair, regardless of code-contributor.}

\tabularnewline

\toprule\noalign{}
& mean & std & min & 25\% & 50\% & 75\% & max \\
\midrule\noalign{}
\endhead
\bottomrule\noalign{}
\endlastfoot
commit\_stats & 0.747 & 0.366 & 0.004 & 0.505 & 1.0 & 1.0 & 1.0 \\
addition\_stats & 0.722 & 0.420 & 0.000 & 0.310 & 1.0 & 1.0 & 1.0 \\
deletion\_stats & 0.737 & 0.417 & 0.000 & 0.722 & 1.0 & 1.0 & 1.0 \\
abs\_stats & 0.733 & 0.404 & 0.000 & 0.417 & 1.0 & 1.0 & 1.0 \\

\end{longtable}

We observed a notable pattern where very few true non-authors (n=4) were
repository owners, while $\sim$49.2\% of unclear cases (n=30)
owned the repositories they contributed to. This suggests that many
unclear contributors were likely primary code authors who could not be
matched due to limited profile information. When excluding repository
owners from the unclear group
(Table~\ref{tbl-unclear-cc-na-commit-stats-no-repo-owner}), the median
contribution drops to $\sim$34.6\% of total commits and
$\sim$12.7\% of absolute changes, though this still represents
substantial technical involvement.

\begin{longtable}[]{@{}llllllll@{}}

\caption{\label{tbl-unclear-cc-na-commit-stats-no-repo-owner}Commit
statistics for code-contributing non-authors labeled as unclear,
excluding repository owners. Statistics are calculated as the proportion
of commits, additions, deletions, and absolute changes made by the
code-contributing non-author to the total commits, additions, deletions,
and absolute changes made to the article-repository pair, regardless of
code-contributor.}

\tabularnewline

\toprule\noalign{}
& mean & std & min & 25\% & 50\% & 75\% & max \\
\midrule\noalign{}
\endhead
\bottomrule\noalign{}
\endlastfoot
commit\_stats & 0.454 & 0.399 & 0.004 & 0.041 & 0.346 & 0.841 & 1.0 \\
addition\_stats & 0.380 & 0.443 & 0.000 & 0.016 & 0.094 & 0.917 & 1.0 \\
deletion\_stats & 0.486 & 0.492 & 0.000 & 0.003 & 0.431 & 0.999 & 1.0 \\
abs\_stats & 0.392 & 0.440 & 0.000 & 0.011 & 0.127 & 0.919 & 1.0 \\

\end{longtable}

Our analysis provides evidence that code-contributing non-authors
represent a heterogeneous group with varying contribution levels. While
defining ``substantial'' contribution worthy of authorship remains
challenging, our findings reveal a clear mix of legitimate non-authors
and potentially missed classifications, with both groups often
contributing meaningful portions of repository commits and code changes.

Our sample size of 200 limits generalizability to the full population of
code-contributing non-authors. Additionally, the manual annotation
process introduces potential subjectivity despite our established
criteria, and our reliance on publicly available GitHub profiles may
systematically underestimate contributions from developers with minimal
profile information.

\subsubsection{Qualitative Error Analysis of Missed
Classifications}\label{sec-appendix-cc-na-error-analysis}

To better understand why certain code-contributing non-authors were
missed classifications, we conducted a qualitative error analysis of the
61 contributors labeled as such. We identified several common themes:

\begin{enumerate}
\def\labelenumi{\arabic{enumi}.}
\tightlist
\item
  \textbf{Limited Information from Text Alone}: The original dataset for
  model training and evaluation was constructed using only text-based
  features from author names and developer information. However, for
  this extended examination, annotators utilized the full
  code-contributor profile, including linked websites or linked ORCID
  profiles. This was done because we wanted to understand the nature of
  missed classifications (with more time and information to make a
  classification) rather than strictly replicating the model's text-only
  approach. From text alone, many of these missed classifications would
  have been very challenging to identify. This highlights a limitation
  in our current model, and a potential area for future work, such as
  incorporating details from linked websites or other contextual
  information to improve matching performance.
\item
  \textbf{Name Variations and Cultural Differences}: The model performed
  better with Anglo-Saxon names, while names from other cultures were
  more likely to be missed. This suggests possible bias in the training
  data and a clear area for future work.
\item
  \textbf{Additional Unrelated Text in Names}: When usernames or display
  names contained longer phrases or unrelated words, the model tended to
  classify them as no-match, even if there were strong indicators of a
  match. For example, a username such as
  ``awesome\_computational\_biologist\_john\_d'' paired with an author
  name ``John Doe'' might be missed due to the additional text in the
  username.
\item
  \textbf{Significant Differences Between Username and Author Name}: The
  model struggled when there were substantial differences between the
  username and author name, such as when an individual provides a chosen
  name in their GitHub profile that differed significantly from their
  authorship name. Most commonly this occurred when an individual used a
  chosen ``English'' name in their GitHub profile that was very
  different from their authorship name.
\end{enumerate}

These themes highlight areas for potential improvement in the model,
such as incorporating more diverse training data and exploring
additional features that could capture cultural name variations and
contextual information.

\subsection{Article Citation Linear Model
Results}\label{article-citation-linear-model-results}

\begin{table}

\caption{\label{tbl-article-composition-overall}Article citations by code contributorship of the research team. Generalized linear model fit with negative binomial distribution and log link function. Significant p-values are indicated with asterisks: p < 0.05 (*), p < 0.01 (**), p < 0.001 (***)}

\centering{

\centering

\begin{tabular}{l*{6}{r}}
\toprule
\textbf{Variable} & \textbf{coef} & \textbf{P>|z|} & \multicolumn{2}{c}{\textbf{[0.025 0.975]}} \\
\midrule
    \bfseries\cellcolor{white}const $^{***}$ & \bfseries\cellcolor{white}0.95 & \bfseries\cellcolor{white}0.00 & \bfseries\cellcolor{white}0.90 & \bfseries\cellcolor{white}1.01 \\
    \bfseries\cellcolor{gray!10}Total Authors $^{***}$ & \bfseries\cellcolor{gray!10}0.08 & \bfseries\cellcolor{gray!10}0.00 & \bfseries\cellcolor{gray!10}0.07 & \bfseries\cellcolor{gray!10}0.09 \\
    \bfseries\cellcolor{white}Code-Contrib. Authors $^{***}$ & \bfseries\cellcolor{white}0.04 & \bfseries\cellcolor{white}0.00 & \bfseries\cellcolor{white}0.02 & \bfseries\cellcolor{white}0.06 \\
    \cellcolor{gray!10}Code-Contrib. Non-Authors & \cellcolor{gray!10}-0.00 & \cellcolor{gray!10}0.61 & \cellcolor{gray!10}-0.01 & \cellcolor{gray!10}0.01 \\
    \bfseries\cellcolor{white}Years Since Publication $^{***}$ & \bfseries\cellcolor{white}0.40 & \bfseries\cellcolor{white}0.00 & \bfseries\cellcolor{white}0.39 & \bfseries\cellcolor{white}0.40 \\
\bottomrule
\end{tabular}

}

\end{table}%

\begin{table}

\caption{\label{tbl-article-composition-oa-status}Article citations by code contributorship of the research team controlled by open access status. Generalized linear model fit with negative binomial distribution and log link function. Significant p-values are indicated with asterisks: p < 0.05 (*), p < 0.01 (**), p < 0.001 (***)}

\centering{

\centering

\begin{tabular}{l*{6}{r}}
\toprule
\textbf{Variable} & \textbf{coef} & \textbf{P>|z|} & \multicolumn{2}{c}{\textbf{[0.025 0.975]}} \\
\midrule
    \bfseries\cellcolor{white}const $^{***}$ & \bfseries\cellcolor{white}0.65 & \bfseries\cellcolor{white}0.00 & \bfseries\cellcolor{white}0.52 & \bfseries\cellcolor{white}0.78 \\
    \bfseries\cellcolor{gray!10}Total Authors $^{***}$ & \bfseries\cellcolor{gray!10}0.08 & \bfseries\cellcolor{gray!10}0.00 & \bfseries\cellcolor{gray!10}0.07 & \bfseries\cellcolor{gray!10}0.09 \\
    \cellcolor{white}Code-Contrib. Authors & \cellcolor{white}-0.03 & \cellcolor{white}0.48 & \cellcolor{white}-0.13 & \cellcolor{white}0.06 \\
    \cellcolor{gray!10}Code-Contrib. Non-Authors & \cellcolor{gray!10}0.01 & \cellcolor{gray!10}0.44 & \cellcolor{gray!10}-0.02 & \cellcolor{gray!10}0.04 \\
    \bfseries\cellcolor{white}Years Since Publication $^{***}$ & \bfseries\cellcolor{white}0.39 & \bfseries\cellcolor{white}0.00 & \bfseries\cellcolor{white}0.38 & \bfseries\cellcolor{white}0.39 \\
    \bfseries\cellcolor{gray!10}Is Open Access $^{***}$ & \bfseries\cellcolor{gray!10}0.34 & \bfseries\cellcolor{gray!10}0.00 & \bfseries\cellcolor{gray!10}0.21 & \bfseries\cellcolor{gray!10}0.46 \\
    \cellcolor{white}Code-Contrib. Authors $\times$ Is Open Access & \cellcolor{white}0.08 & \cellcolor{white}0.12 & \cellcolor{white}-0.02 & \cellcolor{white}0.17 \\
    \cellcolor{gray!10}Code-Contrib. Non-Authors $\times$ Is Open Access & \cellcolor{gray!10}-0.01 & \cellcolor{gray!10}0.39 & \cellcolor{gray!10}-0.05 & \cellcolor{gray!10}0.02 \\
\bottomrule
\end{tabular}

}

\end{table}%

\begin{table}

\caption{\label{tbl-article-composition-domain}Article citations by code contributorship of the research team controlled by domain. Generalized linear model fit with negative binomial distribution and log link function. Significant p-values are indicated with asterisks: p < 0.05 (*), p < 0.01 (**), p < 0.001 (***)}

\centering{

\centering

\begin{tabular}{l*{6}{r}}
\toprule
\textbf{Variable} & \textbf{coef} & \textbf{P>|z|} & \multicolumn{2}{c}{\textbf{[0.025 0.975]}} \\
\midrule
    \bfseries\cellcolor{white}const $^{***}$ & \bfseries\cellcolor{white}0.87 & \bfseries\cellcolor{white}0.00 & \bfseries\cellcolor{white}0.73 & \bfseries\cellcolor{white}1.02 \\
    \bfseries\cellcolor{gray!10}Total Authors $^{***}$ & \bfseries\cellcolor{gray!10}0.08 & \bfseries\cellcolor{gray!10}0.00 & \bfseries\cellcolor{gray!10}0.07 & \bfseries\cellcolor{gray!10}0.09 \\
    \cellcolor{white}Code-Contrib. Authors & \cellcolor{white}-0.00 & \cellcolor{white}0.98 & \cellcolor{white}-0.12 & \cellcolor{white}0.11 \\
    \cellcolor{gray!10}Code-Contrib. Non-Authors & \cellcolor{gray!10}0.02 & \cellcolor{gray!10}0.48 & \cellcolor{gray!10}-0.03 & \cellcolor{gray!10}0.07 \\
    \bfseries\cellcolor{white}Years Since Publication $^{***}$ & \bfseries\cellcolor{white}0.40 & \bfseries\cellcolor{white}0.00 & \bfseries\cellcolor{white}0.39 & \bfseries\cellcolor{white}0.41 \\
    \bfseries\cellcolor{gray!10}Domain Life Sciences $^{**}$ & \bfseries\cellcolor{gray!10}-0.23 & \bfseries\cellcolor{gray!10}0.01 & \bfseries\cellcolor{gray!10}-0.39 & \bfseries\cellcolor{gray!10}-0.06 \\
    \cellcolor{white}Domain Physical Sciences & \cellcolor{white}0.11 & \cellcolor{white}0.13 & \cellcolor{white}-0.03 & \cellcolor{white}0.25 \\
    \bfseries\cellcolor{gray!10}Domain Social Sciences $^{**}$ & \bfseries\cellcolor{gray!10}-0.24 & \bfseries\cellcolor{gray!10}0.01 & \bfseries\cellcolor{gray!10}-0.42 & \bfseries\cellcolor{gray!10}-0.06 \\
    \cellcolor{white}Code-Contrib. Authors $\times$ Domain Life Sciences & \cellcolor{white}0.10 & \cellcolor{white}0.17 & \cellcolor{white}-0.04 & \cellcolor{white}0.23 \\
    \cellcolor{gray!10}Code-Contrib. Authors $\times$ Domain Physical Sciences & \cellcolor{gray!10}0.03 & \cellcolor{gray!10}0.57 & \cellcolor{gray!10}-0.08 & \cellcolor{gray!10}0.15 \\
    \cellcolor{white}Code-Contrib. Authors $\times$ Domain Social Sciences & \cellcolor{white}0.13 & \cellcolor{white}0.07 & \cellcolor{white}-0.01 & \cellcolor{white}0.28 \\
    \cellcolor{gray!10}Code-Contrib. Non-Authors $\times$ Domain Life Sciences & \cellcolor{gray!10}-0.05 & \cellcolor{gray!10}0.19 & \cellcolor{gray!10}-0.12 & \cellcolor{gray!10}0.02 \\
    \cellcolor{white}Code-Contrib. Non-Authors $\times$ Domain Physical Sciences & \cellcolor{white}-0.02 & \cellcolor{white}0.41 & \cellcolor{white}-0.08 & \cellcolor{white}0.03 \\
    \cellcolor{gray!10}Code-Contrib. Non-Authors $\times$ Domain Social Sciences & \cellcolor{gray!10}-0.03 & \cellcolor{gray!10}0.35 & \cellcolor{gray!10}-0.11 & \cellcolor{gray!10}0.04 \\
\bottomrule
\end{tabular}

}

\end{table}%

\begin{table}

\caption{\label{tbl-article-composition-type}Article citations by code contributorship of the research team controlled by article type. Generalized linear model fit with negative binomial distribution and log link function. Significant p-values are indicated with asterisks: p < 0.05 (*), p < 0.01 (**), p < 0.001 (***)}

\centering{

\centering

\begin{tabular}{l*{6}{r}}
\toprule
\textbf{Variable} & \textbf{coef} & \textbf{P>|z|} & \multicolumn{2}{c}{\textbf{[0.025 0.975]}} \\
\midrule
    \bfseries\cellcolor{white}const $^{***}$ & \bfseries\cellcolor{white}0.44 & \bfseries\cellcolor{white}0.00 & \bfseries\cellcolor{white}0.35 & \bfseries\cellcolor{white}0.53 \\
    \bfseries\cellcolor{gray!10}Total Authors $^{***}$ & \bfseries\cellcolor{gray!10}0.08 & \bfseries\cellcolor{gray!10}0.00 & \bfseries\cellcolor{gray!10}0.07 & \bfseries\cellcolor{gray!10}0.09 \\
    \cellcolor{white}Code-Contrib. Authors & \cellcolor{white}-0.01 & \cellcolor{white}0.86 & \cellcolor{white}-0.06 & \cellcolor{white}0.05 \\
    \bfseries\cellcolor{gray!10}Code-Contrib. Non-Authors $^{**}$ & \bfseries\cellcolor{gray!10}-0.03 & \bfseries\cellcolor{gray!10}0.00 & \bfseries\cellcolor{gray!10}-0.05 & \bfseries\cellcolor{gray!10}-0.01 \\
    \bfseries\cellcolor{white}Years Since Publication $^{***}$ & \bfseries\cellcolor{white}0.41 & \bfseries\cellcolor{white}0.00 & \bfseries\cellcolor{white}0.40 & \bfseries\cellcolor{white}0.42 \\
    \bfseries\cellcolor{gray!10}Article Type Research Article $^{***}$ & \bfseries\cellcolor{gray!10}0.53 & \bfseries\cellcolor{gray!10}0.00 & \bfseries\cellcolor{gray!10}0.45 & \bfseries\cellcolor{gray!10}0.61 \\
    \bfseries\cellcolor{white}Article Type Software Article $^{**}$ & \bfseries\cellcolor{white}-0.46 & \bfseries\cellcolor{white}0.00 & \bfseries\cellcolor{white}-0.73 & \bfseries\cellcolor{white}-0.19 \\
    \bfseries\cellcolor{gray!10}Code-Contrib. Authors $\times$ Article Type Research Article $^{*}$ & \bfseries\cellcolor{gray!10}0.07 & \bfseries\cellcolor{gray!10}0.04 & \bfseries\cellcolor{gray!10}0.00 & \bfseries\cellcolor{gray!10}0.13 \\
    \cellcolor{white}Code-Contrib. Authors $\times$ Article Type Software Article & \cellcolor{white}-0.08 & \cellcolor{white}0.23 & \cellcolor{white}-0.22 & \cellcolor{white}0.05 \\
    \bfseries\cellcolor{gray!10}Code-Contrib. Non-Authors $\times$ Article Type Research Article $^{**}$ & \bfseries\cellcolor{gray!10}0.04 & \bfseries\cellcolor{gray!10}0.00 & \bfseries\cellcolor{gray!10}0.02 & \bfseries\cellcolor{gray!10}0.06 \\
    \cellcolor{white}Code-Contrib. Non-Authors $\times$ Article Type Software Article & \cellcolor{white}0.08 & \cellcolor{white}0.28 & \cellcolor{white}-0.07 & \cellcolor{white}0.24 \\
\bottomrule
\end{tabular}

}

\end{table}%

\subsection{Post-Hoc Tests for Coding vs Non-Coding Authors by
Position}\label{post-hoc-tests-for-coding-vs-non-coding-authors-by-position}

\begin{table}

\caption{\label{tbl-post-hoc-tests-on-author-positions}Counts of Code-Contributing Authors ('Coding') and Total Authors by Position and Bonferroni Corrected p-values from Post-Hoc Binomial Tests. Significant p-values are indicated with asterisks: p < 0.05 (*), p < 0.01 (**), p < 0.001 (***).}

\centering{

\centering
\small

\begin{tabular}{llllrr}
\toprule
\textbf{Control} & \textbf{Subset} & \textbf{Position} & \textbf{Coding} & \textbf{Total} & \textbf{p} \\
\midrule
    \multirow{12}{*}{\textbf{Domain}} & \multirow{3}{*}{\textbf{Health Sciences}} & First & 627 & 929 & 0.000$^{***}$ \\
    & & \cellcolor{gray!10}Middle & \cellcolor{gray!10}196 & \cellcolor{gray!10}3488 & \cellcolor{gray!10}0.000$^{***}$ \\
    & & Last & 83 & 920 & 0.000$^{***}$ \\
    & \multirow{3}{*}{\textbf{Life Sciences}} & \cellcolor{gray!10}First & \cellcolor{gray!10}1119 & \cellcolor{gray!10}1684 & \cellcolor{gray!10}0.000$^{***}$ \\
    & & Middle & 383 & 5082 & 0.000$^{***}$ \\
    & & \cellcolor{gray!10}Last & \cellcolor{gray!10}216 & \cellcolor{gray!10}1680 & \cellcolor{gray!10}0.000$^{***}$ \\
    & \multirow{3}{*}{\textbf{Physical Sciences}} & First & 11314 & 16091 & 0.000$^{***}$ \\
    & & \cellcolor{gray!10}Middle & \cellcolor{gray!10}4465 & \cellcolor{gray!10}44332 & \cellcolor{gray!10}0.000$^{***}$ \\
    & & Last & 1074 & 15955 & 0.000$^{***}$ \\
    & \multirow{3}{*}{\textbf{Social Sciences}} & \cellcolor{gray!10}First & \cellcolor{gray!10}786 & \cellcolor{gray!10}1119 & \cellcolor{gray!10}0.000$^{***}$ \\
    & & Middle & 356 & 2770 & 0.000$^{***}$ \\
    & & \cellcolor{gray!10}Last & \cellcolor{gray!10}119 & \cellcolor{gray!10}1118 & \cellcolor{gray!10}0.000$^{***}$ \\
    \midrule
    \multirow{9}{*}{\textbf{Article Type}} & \multirow{3}{*}{\textbf{Preprint}} & First & 1872 & 2662 & 0.000$^{***}$ \\
    & & \cellcolor{gray!10}Middle & \cellcolor{gray!10}791 & \cellcolor{gray!10}7335 & \cellcolor{gray!10}0.000$^{***}$ \\
    & & Last & 174 & 2656 & 0.000$^{***}$ \\
    & \multirow{3}{*}{\textbf{Research Article}} & \cellcolor{gray!10}First & \cellcolor{gray!10}11848 & \cellcolor{gray!10}16961 & \cellcolor{gray!10}0.000$^{***}$ \\
    & & Middle & 4471 & 47814 & 0.000$^{***}$ \\
    & & \cellcolor{gray!10}Last & \cellcolor{gray!10}1276 & \cellcolor{gray!10}16817 & \cellcolor{gray!10}0.000$^{***}$ \\
    & \multirow{3}{*}{\textbf{Software Article}} & First & 126 & 200 & 0.002$^{**}$ \\
    & & \cellcolor{gray!10}Middle & \cellcolor{gray!10}138 & \cellcolor{gray!10}523 & \cellcolor{gray!10}0.000$^{***}$ \\
    & & Last & 42 & 200 & 0.000$^{***}$ \\
    \midrule
    \multirow{6}{*}{\textbf{Open Access Status}} & \multirow{3}{*}{\textbf{Closed Access}} & \cellcolor{gray!10}First & \cellcolor{gray!10}772 & \cellcolor{gray!10}1066 & \cellcolor{gray!10}0.000$^{***}$ \\
    & & Middle & 298 & 3250 & 0.000$^{***}$ \\
    & & \cellcolor{gray!10}Last & \cellcolor{gray!10}70 & \cellcolor{gray!10}1053 & \cellcolor{gray!10}0.000$^{***}$ \\
    & \multirow{3}{*}{\textbf{Open Access}} & First & 13074 & 18757 & 0.000$^{***}$ \\
    & & \cellcolor{gray!10}Middle & \cellcolor{gray!10}5102 & \cellcolor{gray!10}52422 & \cellcolor{gray!10}0.000$^{***}$ \\
    & & Last & 1422 & 18620 & 0.000$^{***}$ \\
    \midrule
    \multirow{3}{*}{\textbf{Overall}} & \multirow{3}{*}{\textbf{Overall}} & \cellcolor{gray!10}First & \cellcolor{gray!10}13846 & \cellcolor{gray!10}19823 & \cellcolor{gray!10}0.000$^{***}$ \\
    & & Middle & 5400 & 55672 & 0.000$^{***}$ \\
    & & \cellcolor{gray!10}Last & \cellcolor{gray!10}1492 & \cellcolor{gray!10}19673 & \cellcolor{gray!10}0.000$^{***}$ \\
\bottomrule
\end{tabular}

}

\end{table}%

Counts of authors in Table~\ref{tbl-post-hoc-tests-on-author-positions}
may differ slightly from counts in
Table~\ref{tbl-team-comp-no-push-after-pub-counts}.
Table~\ref{tbl-team-comp-no-push-after-pub-counts} counts unique
authors, while Table~\ref{tbl-post-hoc-tests-on-author-positions} counts
unique author-document pairs (i.e., the same author may appear in
multiple documents).

\subsection{Post-Hoc Tests for Coding vs Non-Coding Authors by
Corresponding
Status}\label{post-hoc-tests-for-coding-vs-non-coding-authors-by-corresponding-status}

\begin{table}

\caption{\label{tbl-post-hoc-tests-on-corresponding-status}Counts of Code-Contributing Authors ('Coding') and Total Authors by Corresponding Status and Bonferroni Corrected p-values from Post-Hoc Binomial Tests. Significant p-values are indicated with asterisks: p < 0.05 (*), p < 0.01 (**), p < 0.001 (***).}

\centering{

\centering
\small

\begin{tabular}{llllrr}
\toprule
\textbf{Control} & \textbf{Subset} & \textbf{Is Corresponding} & \textbf{Coding} & \textbf{Total} & \textbf{p} \\
\midrule
    \multirow{8}{*}{\textbf{Domain}} & \multirow{2}{*}{\textbf{Health Sciences}} & Corresponding & 407 & 1954 & 0.000$^{***}$ \\
    & & \cellcolor{gray!10}Not Corresponding & \cellcolor{gray!10}499 & \cellcolor{gray!10}3383 & \cellcolor{gray!10}0.000$^{***}$ \\
    & \multirow{2}{*}{\textbf{Life Sciences}} & Corresponding & 871 & 3975 & 0.000$^{***}$ \\
    & & \cellcolor{gray!10}Not Corresponding & \cellcolor{gray!10}847 & \cellcolor{gray!10}4471 & \cellcolor{gray!10}0.000$^{***}$ \\
    & \multirow{2}{*}{\textbf{Physical Sciences}} & Corresponding & 2213 & 5972 & 0.000$^{***}$ \\
    & & \cellcolor{gray!10}Not Corresponding & \cellcolor{gray!10}14640 & \cellcolor{gray!10}70406 & \cellcolor{gray!10}0.000$^{***}$ \\
    & \multirow{2}{*}{\textbf{Social Sciences}} & Corresponding & 330 & 959 & 0.000$^{***}$ \\
    & & \cellcolor{gray!10}Not Corresponding & \cellcolor{gray!10}931 & \cellcolor{gray!10}4048 & \cellcolor{gray!10}0.000$^{***}$ \\
    \midrule
    \multirow{6}{*}{\textbf{Article Type}} & \multirow{2}{*}{\textbf{Preprint}} & Corresponding & 13 & 41 & 0.055 \\
    & & \cellcolor{gray!10}Not Corresponding & \cellcolor{gray!10}2824 & \cellcolor{gray!10}12612 & \cellcolor{gray!10}0.000$^{***}$ \\
    & \multirow{2}{*}{\textbf{Research Article}} & Corresponding & 3739 & 12638 & 0.000$^{***}$ \\
    & & \cellcolor{gray!10}Not Corresponding & \cellcolor{gray!10}13856 & \cellcolor{gray!10}68954 & \cellcolor{gray!10}0.000$^{***}$ \\
    & \multirow{2}{*}{\textbf{Software Article}} & Corresponding & 69 & 181 & 0.003$^{**}$ \\
    & & \cellcolor{gray!10}Not Corresponding & \cellcolor{gray!10}237 & \cellcolor{gray!10}742 & \cellcolor{gray!10}0.000$^{***}$ \\
    \midrule
    \multirow{4}{*}{\textbf{Open Access Status}} & \multirow{2}{*}{\textbf{Closed Access}} & Corresponding & 82 & 181 & 0.468 \\
    & & \cellcolor{gray!10}Not Corresponding & \cellcolor{gray!10}1058 & \cellcolor{gray!10}5188 & \cellcolor{gray!10}0.000$^{***}$ \\
    & \multirow{2}{*}{\textbf{Open Access}} & Corresponding & 3739 & 12679 & 0.000$^{***}$ \\
    & & \cellcolor{gray!10}Not Corresponding & \cellcolor{gray!10}15859 & \cellcolor{gray!10}77120 & \cellcolor{gray!10}0.000$^{***}$ \\
    \midrule
    \multirow{2}{*}{\textbf{Overall}} & \multirow{2}{*}{\textbf{Overall}} & Corresponding & 3821 & 12860 & 0.000$^{***}$ \\
    & & \cellcolor{gray!10}Not Corresponding & \cellcolor{gray!10}16917 & \cellcolor{gray!10}82308 & \cellcolor{gray!10}0.000$^{***}$ \\
\bottomrule
\end{tabular}

}

\end{table}%

Counts of authors in
Table~\ref{tbl-post-hoc-tests-on-corresponding-status} may differ
slightly from counts in
Table~\ref{tbl-team-comp-no-push-after-pub-counts}.
Table~\ref{tbl-team-comp-no-push-after-pub-counts} counts unique
authors, while Table~\ref{tbl-post-hoc-tests-on-corresponding-status}
counts unique author-document pairs (i.e., the same author may appear in
multiple documents).

\subsection{Filtered Dataset Description for h-Index
Analysis}\label{filtered-dataset-description-for-h-index-analysis}

\begin{table}

\caption{\label{tbl-h-index-counts}Counts of Total Authors, n Any Coding Authors, n Majority Coding Authors, and n Always Coding Authors by Common Domain, Document Type, and Author Position. Authors are only included if they have three or more publications within our dataset and are associated with no more than three developer accounts, with each association having a predicted model confidence of at least 97\%.}

\centering{

\centering
\small

\begin{tabular}{llrrrr}
\toprule
\textbf{Category} & \textbf{Subset} & \textbf{Total Authors} & \textbf{Any Code} & \textbf{Majority Code} & \textbf{Always Code} \\
\midrule
    \multirow{4}{*}{\textbf{By Commmon Domain}} & \cellcolor{gray!10}Health Sciences & \cellcolor{gray!10}1501 & \cellcolor{gray!10}338 & \cellcolor{gray!10}196 & \cellcolor{gray!10}81 \\
    & Life Sciences & 1436 & 351 & 236 & 127 \\
    & \cellcolor{gray!10}Physical Sciences & \cellcolor{gray!10}49447 & \cellcolor{gray!10}14756 & \cellcolor{gray!10}7954 & \cellcolor{gray!10}3725 \\
    & Social Sciences & 1297 & 274 & 216 & 176 \\\midrule
    \multirow{3}{*}{\textbf{By Document Type}} & \cellcolor{gray!10}Preprint & \cellcolor{gray!10}29038 & \cellcolor{gray!10}9255 & \cellcolor{gray!10}4828 & \cellcolor{gray!10}2151 \\
    & research article & 24265 & 6419 & 3657 & 1830 \\
    & \cellcolor{gray!10}software article & \cellcolor{gray!10}378 & \cellcolor{gray!10}45 & \cellcolor{gray!10}117 & \cellcolor{gray!10}128 \\\midrule
    \multirow{3}{*}{\textbf{By Author Position}} & \cellcolor{gray!10}First & \cellcolor{gray!10}11459 & \cellcolor{gray!10}1671 & \cellcolor{gray!10}4864 & \cellcolor{gray!10}3249 \\
    & last & 10208 & 2260 & 550 & 186 \\
    & \cellcolor{gray!10}middle & \cellcolor{gray!10}32014 & \cellcolor{gray!10}11788 & \cellcolor{gray!10}3188 & \cellcolor{gray!10}674 \\\midrule
    \textbf{Total} & & \textbf{53681} & \textbf{15719} & \textbf{8602} & \textbf{4109} \\
\bottomrule
\end{tabular}

}

\end{table}%

\subsection{h-Index Linear Model
Results}\label{h-index-linear-model-results}

\begin{table}

\caption{\label{tbl-researcher-coding-status-no-control}Code-contributing authors h-index by coding status. Generalized linear model fit with Gaussian distribution and log link function. Significant p-values are indicated with asterisks: p < 0.05 (*), p < 0.01 (**), p < 0.001 (***).}

\centering{

\centering

\begin{tabular}{l*{6}{r}}
\toprule
\textbf{Variable} & \textbf{coef} & \textbf{P>|z|} & \multicolumn{2}{c}{\textbf{[0.025 0.975]}} \\
\midrule
    \bfseries\cellcolor{white}const $^{***}$ & \bfseries\cellcolor{white}3.18 & \bfseries\cellcolor{white}0.00 & \bfseries\cellcolor{white}3.17 & \bfseries\cellcolor{white}3.19 \\
    \bfseries\cellcolor{gray!10}Works Count $^{***}$ & \bfseries\cellcolor{gray!10}0.00 & \bfseries\cellcolor{gray!10}0.00 & \bfseries\cellcolor{gray!10}0.00 & \bfseries\cellcolor{gray!10}0.00 \\
    \bfseries\cellcolor{white}Any Coding $^{***}$ & \bfseries\cellcolor{white}-0.32 & \bfseries\cellcolor{white}0.00 & \bfseries\cellcolor{white}-0.34 & \bfseries\cellcolor{white}-0.31 \\
    \bfseries\cellcolor{gray!10}Majority Coding $^{***}$ & \bfseries\cellcolor{gray!10}-0.76 & \bfseries\cellcolor{gray!10}0.00 & \bfseries\cellcolor{gray!10}-0.79 & \bfseries\cellcolor{gray!10}-0.73 \\
    \bfseries\cellcolor{white}Always Coding $^{***}$ & \bfseries\cellcolor{white}-0.96 & \bfseries\cellcolor{white}0.00 & \bfseries\cellcolor{white}-1.01 & \bfseries\cellcolor{white}-0.91 \\
\bottomrule
\end{tabular}

}

\end{table}%

\begin{table}

\caption{\label{tbl-researcher-coding-status-author-position}Code-contributing authors h-index by coding status controlled by most freq. author position. Generalized linear model fit with Gaussian distribution and log link function. Significant p-values are indicated with asterisks: p < 0.05 (*), p < 0.01 (**), p < 0.001 (***).}

\centering{

\centering

\begin{tabular}{l*{6}{r}}
\toprule
\textbf{Variable} & \textbf{coef} & \textbf{P>|z|} & \multicolumn{2}{c}{\textbf{[0.025 0.975]}} \\
\midrule
    \bfseries\cellcolor{white}const $^{***}$ & \bfseries\cellcolor{white}2.36 & \bfseries\cellcolor{white}0.00 & \bfseries\cellcolor{white}2.30 & \bfseries\cellcolor{white}2.42 \\
    \bfseries\cellcolor{gray!10}Works Count $^{***}$ & \bfseries\cellcolor{gray!10}0.00 & \bfseries\cellcolor{gray!10}0.00 & \bfseries\cellcolor{gray!10}0.00 & \bfseries\cellcolor{gray!10}0.00 \\
    \bfseries\cellcolor{white}Any Coding $^{***}$ & \bfseries\cellcolor{white}0.16 & \bfseries\cellcolor{white}0.00 & \bfseries\cellcolor{white}0.07 & \bfseries\cellcolor{white}0.24 \\
    \cellcolor{gray!10}Majority Coding & \cellcolor{gray!10}-0.05 & \cellcolor{gray!10}0.19 & \cellcolor{gray!10}-0.12 & \cellcolor{gray!10}0.03 \\
    \bfseries\cellcolor{white}Always Coding $^{***}$ & \bfseries\cellcolor{white}-0.22 & \bfseries\cellcolor{white}0.00 & \bfseries\cellcolor{white}-0.31 & \bfseries\cellcolor{white}-0.14 \\
    \bfseries\cellcolor{gray!10}Common Author Position Last $^{***}$ & \bfseries\cellcolor{gray!10}1.06 & \bfseries\cellcolor{gray!10}0.00 & \bfseries\cellcolor{gray!10}1.00 & \bfseries\cellcolor{gray!10}1.13 \\
    \bfseries\cellcolor{white}Common Author Position Middle $^{***}$ & \bfseries\cellcolor{white}0.75 & \bfseries\cellcolor{white}0.00 & \bfseries\cellcolor{white}0.69 & \bfseries\cellcolor{white}0.82 \\
    \bfseries\cellcolor{gray!10}Any Coding $\times$ Common Author Position Last $^{***}$ & \bfseries\cellcolor{gray!10}-0.31 & \bfseries\cellcolor{gray!10}0.00 & \bfseries\cellcolor{gray!10}-0.40 & \bfseries\cellcolor{gray!10}-0.23 \\
    \bfseries\cellcolor{white}Any Coding $\times$ Common Author Position Middle $^{***}$ & \bfseries\cellcolor{white}-0.47 & \bfseries\cellcolor{white}0.00 & \bfseries\cellcolor{white}-0.55 & \bfseries\cellcolor{white}-0.38 \\
    \bfseries\cellcolor{gray!10}Majority Coding $\times$ Common Author Position Last $^{***}$ & \bfseries\cellcolor{gray!10}-0.37 & \bfseries\cellcolor{gray!10}0.00 & \bfseries\cellcolor{gray!10}-0.47 & \bfseries\cellcolor{gray!10}-0.28 \\
    \bfseries\cellcolor{white}Majority Coding $\times$ Common Author Position Middle $^{***}$ & \bfseries\cellcolor{white}-0.62 & \bfseries\cellcolor{white}0.00 & \bfseries\cellcolor{white}-0.71 & \bfseries\cellcolor{white}-0.54 \\
    \bfseries\cellcolor{gray!10}Always Coding $\times$ Common Author Position Last $^{***}$ & \bfseries\cellcolor{gray!10}-0.38 & \bfseries\cellcolor{gray!10}0.00 & \bfseries\cellcolor{gray!10}-0.52 & \bfseries\cellcolor{gray!10}-0.23 \\
    \bfseries\cellcolor{white}Always Coding $\times$ Common Author Position Middle $^{***}$ & \bfseries\cellcolor{white}-0.51 & \bfseries\cellcolor{white}0.00 & \bfseries\cellcolor{white}-0.64 & \bfseries\cellcolor{white}-0.38 \\
\bottomrule
\end{tabular}

}

\end{table}%

\begin{table}

\caption{\label{tbl-researcher-coding-status-domain}Code-contributing authors h-index by coding status controlled by most freq. domain. Generalized linear model fit with Gaussian distribution and log link function. Significant p-values are indicated with asterisks: p < 0.05 (*), p < 0.01 (**), p < 0.001 (***).}

\centering{

\centering

\begin{tabular}{l*{6}{r}}
\toprule
\textbf{Variable} & \textbf{coef} & \textbf{P>|z|} & \multicolumn{2}{c}{\textbf{[0.025 0.975]}} \\
\midrule
    \bfseries\cellcolor{white}const $^{***}$ & \bfseries\cellcolor{white}3.31 & \bfseries\cellcolor{white}0.00 & \bfseries\cellcolor{white}3.28 & \bfseries\cellcolor{white}3.35 \\
    \bfseries\cellcolor{gray!10}Works Count $^{***}$ & \bfseries\cellcolor{gray!10}0.00 & \bfseries\cellcolor{gray!10}0.00 & \bfseries\cellcolor{gray!10}0.00 & \bfseries\cellcolor{gray!10}0.00 \\
    \bfseries\cellcolor{white}Any Coding $^{***}$ & \bfseries\cellcolor{white}-0.38 & \bfseries\cellcolor{white}0.00 & \bfseries\cellcolor{white}-0.47 & \bfseries\cellcolor{white}-0.29 \\
    \bfseries\cellcolor{gray!10}Majority Coding $^{***}$ & \bfseries\cellcolor{gray!10}-1.47 & \bfseries\cellcolor{gray!10}0.00 & \bfseries\cellcolor{gray!10}-1.67 & \bfseries\cellcolor{gray!10}-1.26 \\
    \bfseries\cellcolor{white}Always Coding $^{***}$ & \bfseries\cellcolor{white}-1.18 & \bfseries\cellcolor{white}0.00 & \bfseries\cellcolor{white}-1.55 & \bfseries\cellcolor{white}-0.82 \\
    \bfseries\cellcolor{gray!10}Common Domain Life Sciences $^{***}$ & \bfseries\cellcolor{gray!10}0.10 & \bfseries\cellcolor{gray!10}0.00 & \bfseries\cellcolor{gray!10}0.06 & \bfseries\cellcolor{gray!10}0.15 \\
    \bfseries\cellcolor{white}Common Domain Physical Sciences $^{***}$ & \bfseries\cellcolor{white}-0.14 & \bfseries\cellcolor{white}0.00 & \bfseries\cellcolor{white}-0.18 & \bfseries\cellcolor{white}-0.11 \\
    \bfseries\cellcolor{gray!10}Common Domain Social Sciences $^{***}$ & \bfseries\cellcolor{gray!10}-0.17 & \bfseries\cellcolor{gray!10}0.00 & \bfseries\cellcolor{gray!10}-0.23 & \bfseries\cellcolor{gray!10}-0.11 \\
    \cellcolor{white}Any Coding $\times$ Common Domain Life Sciences & \cellcolor{white}0.08 & \cellcolor{white}0.16 & \cellcolor{white}-0.03 & \cellcolor{white}0.20 \\
    \cellcolor{gray!10}Any Coding $\times$ Common Domain Physical Sciences & \cellcolor{gray!10}0.06 & \cellcolor{gray!10}0.17 & \cellcolor{gray!10}-0.03 & \cellcolor{gray!10}0.15 \\
    \cellcolor{white}Any Coding $\times$ Common Domain Social Sciences & \cellcolor{white}0.02 & \cellcolor{white}0.81 & \cellcolor{white}-0.13 & \cellcolor{white}0.16 \\
    \bfseries\cellcolor{gray!10}Majority Coding $\times$ Common Domain Life Sciences $^{***}$ & \bfseries\cellcolor{gray!10}0.84 & \bfseries\cellcolor{gray!10}0.00 & \bfseries\cellcolor{gray!10}0.61 & \bfseries\cellcolor{gray!10}1.08 \\
    \bfseries\cellcolor{white}Majority Coding $\times$ Common Domain Physical Sciences $^{***}$ & \bfseries\cellcolor{white}0.73 & \bfseries\cellcolor{white}0.00 & \bfseries\cellcolor{white}0.52 & \bfseries\cellcolor{white}0.93 \\
    \bfseries\cellcolor{gray!10}Majority Coding $\times$ Common Domain Social Sciences $^{***}$ & \bfseries\cellcolor{gray!10}0.75 & \bfseries\cellcolor{gray!10}0.00 & \bfseries\cellcolor{gray!10}0.48 & \bfseries\cellcolor{gray!10}1.02 \\
    \cellcolor{white}Always Coding $\times$ Common Domain Life Sciences & \cellcolor{white}0.28 & \cellcolor{white}0.19 & \cellcolor{white}-0.14 & \cellcolor{white}0.70 \\
    \cellcolor{gray!10}Always Coding $\times$ Common Domain Physical Sciences & \cellcolor{gray!10}0.23 & \cellcolor{gray!10}0.23 & \cellcolor{gray!10}-0.14 & \cellcolor{gray!10}0.59 \\
    \cellcolor{white}Always Coding $\times$ Common Domain Social Sciences & \cellcolor{white}0.28 & \cellcolor{white}0.20 & \cellcolor{white}-0.15 & \cellcolor{white}0.72 \\
\bottomrule
\end{tabular}

}

\end{table}%

\begin{table}

\caption{\label{tbl-researcher-coding-status-article-type}Code-contributing authors h-index by coding status controlled by most freq. article type. Generalized linear model fit with Gaussian distribution and log link function. Significant p-values are indicated with asterisks: p < 0.05 (*), p < 0.01 (**), p < 0.001 (***).}

\centering{

\centering

\begin{tabular}{l*{6}{r}}
\toprule
\textbf{Variable} & \textbf{coef} & \textbf{P>|z|} & \multicolumn{2}{c}{\textbf{[0.025 0.975]}} \\
\midrule
    \bfseries\cellcolor{white}const $^{***}$ & \bfseries\cellcolor{white}3.09 & \bfseries\cellcolor{white}0.00 & \bfseries\cellcolor{white}3.08 & \bfseries\cellcolor{white}3.10 \\
    \bfseries\cellcolor{gray!10}Works Count $^{***}$ & \bfseries\cellcolor{gray!10}0.00 & \bfseries\cellcolor{gray!10}0.00 & \bfseries\cellcolor{gray!10}0.00 & \bfseries\cellcolor{gray!10}0.00 \\
    \bfseries\cellcolor{white}Any Coding $^{***}$ & \bfseries\cellcolor{white}-0.29 & \bfseries\cellcolor{white}0.00 & \bfseries\cellcolor{white}-0.32 & \bfseries\cellcolor{white}-0.27 \\
    \bfseries\cellcolor{gray!10}Majority Coding $^{***}$ & \bfseries\cellcolor{gray!10}-0.76 & \bfseries\cellcolor{gray!10}0.00 & \bfseries\cellcolor{gray!10}-0.80 & \bfseries\cellcolor{gray!10}-0.72 \\
    \bfseries\cellcolor{white}Always Coding $^{***}$ & \bfseries\cellcolor{white}-0.98 & \bfseries\cellcolor{white}0.00 & \bfseries\cellcolor{white}-1.06 & \bfseries\cellcolor{white}-0.90 \\
    \bfseries\cellcolor{gray!10}Common Article Type Research Article $^{***}$ & \bfseries\cellcolor{gray!10}0.18 & \bfseries\cellcolor{gray!10}0.00 & \bfseries\cellcolor{gray!10}0.17 & \bfseries\cellcolor{gray!10}0.20 \\
    \bfseries\cellcolor{white}Common Article Type Software Article $^{***}$ & \bfseries\cellcolor{white}0.22 & \bfseries\cellcolor{white}0.00 & \bfseries\cellcolor{white}0.12 & \bfseries\cellcolor{white}0.33 \\
    \bfseries\cellcolor{gray!10}Any Coding $\times$ Common Article Type Research Article $^{*}$ & \bfseries\cellcolor{gray!10}-0.03 & \bfseries\cellcolor{gray!10}0.05 & \bfseries\cellcolor{gray!10}-0.06 & \bfseries\cellcolor{gray!10}-0.00 \\
    \cellcolor{white}Any Coding $\times$ Common Article Type Software Article & \cellcolor{white}0.17 & \cellcolor{white}0.10 & \cellcolor{white}-0.04 & \cellcolor{white}0.37 \\
    \cellcolor{gray!10}Majority Coding $\times$ Common Article Type Research Article & \cellcolor{gray!10}0.01 & \cellcolor{gray!10}0.80 & \cellcolor{gray!10}-0.05 & \cellcolor{gray!10}0.06 \\
    \bfseries\cellcolor{white}Majority Coding $\times$ Common Article Type Software Article $^{***}$ & \bfseries\cellcolor{white}0.38 & \bfseries\cellcolor{white}0.00 & \bfseries\cellcolor{white}0.20 & \bfseries\cellcolor{white}0.56 \\
    \cellcolor{gray!10}Always Coding $\times$ Common Article Type Research Article & \cellcolor{gray!10}0.00 & \cellcolor{gray!10}0.97 & \cellcolor{gray!10}-0.10 & \cellcolor{gray!10}0.10 \\
    \bfseries\cellcolor{white}Always Coding $\times$ Common Article Type Software Article $^{**}$ & \bfseries\cellcolor{white}0.37 & \bfseries\cellcolor{white}0.00 & \bfseries\cellcolor{white}0.16 & \bfseries\cellcolor{white}0.58 \\
\bottomrule
\end{tabular}

}

\end{table}%

\subsection{Study Differences from
Preregistration}\label{study-differences-from-preregistration}

\subsubsection{Analysis of Article Field Weighted Citation Impact (FWCI)
and Code
Contribution}\label{analysis-of-article-field-weighted-citation-impact-fwci-and-code-contribution}

In our pre-registered analysis plan (\url{https://osf.io/fc74m}), we
initially stated that we would additionally investigate the relationship
between an article's Field Weighted Citation Impact (FWCI) and the
number of code contributors to the project. We decided against this
analysis as the FWCI metric was only available from OpenAlex for 55.5\%
(n=76904) articles from the \texttt{rs-graph-v1} dataset at the time of
data processing. In addition, our analysis of the relationship between
article citations and the number of code contributors to the project
already includes the articles domain and duration since publication
providing similar control.

\subsubsection{Analysis of Project Duration and Percentage
Code-Contributors Who Are
Authors}\label{analysis-of-project-duration-and-percentage-code-contributors-who-are-authors}

In our pre-registered analysis plan (\url{https://osf.io/fc74m}), we
initially hypothesized that there would be a positive relationship
between project duration and authorship recognition. Specifically, we
posited that sustained technical engagement and scientific recognition
might be meaningfully related, with longer project durations potentially
leading to higher rates of code-contributor authorship. We saw
repository histories as providing a unique opportunity to examine this
relationship, leading us to hypothesize that projects with longer commit
durations would be associated with higher percentages of developers
receiving authorship recognition (pre-registered as H2).

However, our analysis found no evidence to support this hypothesis. When
examining the relationship between a repository's commit duration and
the percentage of developers who receive authorship recognition, we
found no significant correlation (r = -0.00, p = n.s.). This suggests
that the length of time a project has been in development has no
meaningful relationship with the proportion of developers who are
recognized as authors.

We ultimately decided to exclude this analysis for two key
methodological reasons. First, our approach of using repository-level
commit duration as a proxy for individual contribution patterns proved
too coarse-grained. A more precise analysis would need to examine
individual-level contribution durations and patterns rather than overall
project length. Second, our method did not account for the varying
levels of contribution that different developers make to a repository.
Simply correlating overall project duration with authorship rates fails
to capture the nuanced ways that sustained, meaningful technical
contributions might influence authorship decisions.

These limitations suggest potential directions for future work that
could more rigorously examine the relationship between long-term
technical engagement and scientific recognition. Such work could benefit
from a more granular analysis of individual contribution patterns,
incorporating measures of contribution significance and sustainability
rather than just temporal duration.

\end{document}